\documentclass{aa}
\usepackage{subfigure}
\usepackage{graphicx}
\usepackage{natbib}
\bibpunct{(}{)}{;}{a}{}{,}
\usepackage{txfonts}
\begin{document}
   \title{High mass star formation in the IRAS 17233$-$3606 region: a new nearby and bright hot core in the southern sky}

   \subtitle{}

  \author{S. Leurini\inst{1,2} \and C. Hieret\inst{2} \and  S. {Thorwirth}\inst{2} \and F. Wyrowski\inst{2}
   \and P. Schilke\inst{2} \and K.M. Menten\inst{2} \and R. G\"usten\inst{2} \and L. Zapata\inst{2}}

   \offprints{S. Leurini}

   \institute{ESO, Karl-Scharzschild-Strasse 2, D-85748,
Garching-bei-M\"unchen\\ \email{sleurini@eso.org}
          \and Max Planck Institut f\"ur Radioastronomie,  Auf dem H\"ugel 69, D-53121, Bonn\\ \email{name@mpifr-bonn.mpg.de}
              }

   \date{\today}

 \abstract{}{We present molecular line observations of the massive
 star forming region IRAS 17233-3606 aimed at studying the molecular
 core associated with the source.}
{The observations were made using
 the Atacama Pathfinder Experiment telescope in the CO (3-2)
 and HCO$^+$ (4-3) transitions, and in the CH$_3$OH (6$_K-5_K$), (7$_K-6_K$)
 and CH$_3$CN (16$_K-15_K$) bands. For the CO(3-2) and HCO$^+$ (4-3)
 transitions, we obtained maps with a size of $70''\times 70''$. The typical angular resolution of the
 data is $\sim 18''$.}
 {Our observations reveal an exceptionally rich molecular spectrum, a
 signpost of hot core activity. Comparisons with two other prominent
 southern hot cores were made through observations in the same
 frequency setups. We also detected a bipolar
 outflow in CO (3-2) and HCO$^+$ (4-3) lines.  Modelling reveals a hot core of size $\sim 3''$ and
a temperature of 150~K in the IRAS\,17233$-$3606 region. The parameters of the molecular
 outflow are derived through the analysis of the CO (3-2) emission, and are typical
of outflows driven by high-mass young stellar objects.}
 {}

   \titlerunning{Star formation in the IRAS 17233$-$3606 region}\authorrunning{Leurini et al.}
   \maketitle
%

\section{Introduction}\label{intro}
The last decade has seen significant progress in the understanding of
how high-mass stars form. Large samples of massive young stellar
objects (YSOs) were studied with single-dish telescopes to
investigate their physical properties through the analysis of their
(sub)mm continuum and molecular emission
\citep[e.g.,][]{1996A&A...308..573M,1998A&A...336..339M,2000A&A...355..617M,1997MNRAS.291..261W,1998MNRAS.301..640W,1999MNRAS.309..905W,2000A&A...357..637H,2001ApJ...552L.167Z,2005ApJ...625..864Z,2002ApJ...566..931S,2002ApJ...566..945B,2002A&A...383..892B,2004A&A...426...97F,2004A&A...417..115W,2005A&A...434..257W}. However, an intrinsic feature of high-mass stars is that they form in clusters,
and that most of them are at large (several kpc) distances. Therefore,
single-dish studies lack the necessary spatial resolution to resolve
single protostars. However, they are essential to perform statistical studies
on large samples of objects, and to identify the best candidates for follow up analysis
at higher resolution.
Since  the Atacama Large Millimeter Array will regularly
operate at 0.1$''$ scales in
the near future, it is now important to select massive
young stellar object candidates that are observable from the
southern hemisphere and relatively close-by in order to reach
linear resolution sufficient for detailed studies of the chemical and physical processes
in the environment of the newly formed high mass stars and the latter's interaction with
their environment.

The region harbouring the prominent  far-infrared source IRAS\,17233$-$3606 (hereafter IRAS\,17233)
first came to attention through its very intense H$_2$O, OH, and CH$_3$OH masers. The combination of the
latter is an unambiguous signpost of recent high-mass star formation \citep[e.g.,][]{1989A&A...213..339F,1998MNRAS.301..640W}.
IRAS 17233, also known as G351.78$-$0.54,  contains  two cm continuum sources with a separation of
$\sim 12''$, one being a compact H{\sc II} region, the other a
weak double UCH{\sc II} source \citep{1993AJ....105.1495H} close to the location of the maser spots.
Recently, higher angular resolution multi-radio wavelengths observations by \citet{zapata}  found a cluster of
four compact radio objects of sizes equal or less than 0.3$''$ and varying (mostly thermal) spectral indices in the region of the double radio source.  One of them is at the centre
of a bipolar outflow traced by OH masers \citep{2005ApJS..160..220F}.
Moreover, non-Gaussian profiles were detected in CO(2-1) by
\citet{1997ApJS..110...71O} which are probably associated with a
molecular outflow. From these data however, the powering source of the
molecular outflow remains unidentified.

Previous
studies seem to agree that IRAS 17233 is located at the near kinematic distance
\citep[between 700~pc and
2.2~kpc,][]{2006A&A...460..721M,1989A&A...213..339F} rather than at
the far distance \citep[$\sim 16$~kpc, e.g. ][]{1998AJ....116.1897M}.
This is suggested by the high measured intensities of practically all emissions, which would indicate exceedingly high luminosities
if the source were at the far distance and, persuasively, by the fact that it is at an
angular distance of more than 0.5 degrees below the Galactic plane.
In the following discussion, we adopt a distance of 1~kpc for the source.
Continuum observations of dust emission at 1.2~mm
\citep{2004A&A...426...97F} imply a mass of 200~M$_\odot$, and a
bolometric luminosity of $2.7~10^4$~L$_\odot$ for the region
follows from its spectral energy distribution.
So far, little data are available on \textit{thermally excited} molecular line emission from IRAS 17233,	
which would allow											
investigation of the physical and chemical status of the region and lay the groundwork for future	
detailed studies with ALMA.

In this paper, we report observations of IRAS\,17233 with
the Atacama Pathfinder Experiment telescope\footnote{This
publication is based on data acquired with the Atacama Pathfinder
Experiment (APEX). APEX is a collaboration between the
Max-Planck-Institut f\"ur Radioastronomie, the European Southern
Observatory, and the Onsala Space Observatory} in several molecular tracers,
 with typical angular
resolutions of 18$''$. These data reveal a bipolar outflow originating
from a position very close to the methanol maser spots detected
by \citet{1998MNRAS.301..640W}, and a rich molecular spectrum
typical of hot cores near massive young stellar objects.

\section{Observations}\label{par-obs}
The observations were made during 2006 April and
August  with the Atacama Pathfinder Experiment 12~m telescope (APEX) located
on Llano de Chajnantor in the Atacama desert of Chile. These
observations are part of a larger program aimed at studying massive
young stellar objects (YSOs) in the southern hemisphere \citep{2007arXiv0706.1643H}.

IRAS\,17233
was observed with the APEX-2A facility receiver
\citep{2006A&A...454L..17R} tuned at 356.7~GHz in the upper side band, in
order to simultaneously detect the HCO$^+$ (4-3) transition (and, from mapping,
retrieve the peak position) as well as  the CO (3-2) line from the lower
side band (to detect any outflow activity).   A map of 70$\arcsec
\times 70 \arcsec$ with a spacing of 10$\arcsec$ was taken centred on the
1.2~mm continuum peak position from
\citet{2004A&A...426...97F}  ($\alpha_{2000}=17^h26^m43.0^s,~\delta_{2000}=
-36^\circ09^\prime 15.5^{\prime\prime}$).
These observations were followed up by ON-OFF integrations on the peak
of the HCO$^+$ (4-3) emission ($\alpha_{2000}=17^h26^m42.2^s,~\delta_{2000}=
-36^\circ09^\prime 21.5^{\prime\prime}$) in CH$_3$OH ($6_K-5_K, v_t=0,1$) at
289.8~GHz in LSB, CH$_3$CN ($16_K-15_K$) at 294.45~GHz in USB, and
CH$_3$OH ($7_K-6_K, v_t=0,1$) at 337.8~GHz in LSB. For comparison, we also
 observed the
massive star forming regions G327.3-0.6
($\alpha_{2000}=15^h53^m08.6^s,~\delta_{2000}= -54^\circ 37^\prime
07.3^{\prime\prime}$) and NGC6334(I)
($\alpha_{2000}=17^h20^m53.4^s,~\delta_{2000}= -35^\circ 47^\prime
01.5^{\prime\prime}$) at 289.8 and 294.45~GHz. The OFF positions were, in all cases, at
(600$''$,0$''$) from the reference positions.
The APEX facility FFT
spectrometers \citep{2006A&A...454L..29K} were used in series with 100~MHz of overlap, thereby
providing 1.8 GHz of bandwidth. Due to the high dynamic range of the
FFTS (48~dB of the 8 bit analog-to-digital converter) the IF processor
had to limit the band edges with steep band pass filters to avoid
aliasing by out-of-band signals. However, because of the limited
steepness of any analog filter, residual aliasing is unavoidable,
thereby limiting the effective bandwidth of the spectrometer.  The region
of the spectrum which can be affected by this problem is limited to the outer $\sim$50~MHz of bands.

In August 2006, continuum cross-scans were performed with the APEX-2A and the FLASH receivers \citep{2006A&A...454L..21H}
at 344.7, 461 and 806.7~GHz respectively,  towards the  1.2~mm continuum peak position from
\citet{2004A&A...426...97F}.

The pointing of the telescope was checked on Jupiter, NGC6334(I) and SGRB2-N and
found to be accurate to 2$''$(rms).

The observed transitions and basic observational parameters are
summarised in Tables~\ref{obs} and \ref{obs-con}. We used a main-beam efficiency of 0.73, 0.60 and 0.43
for APEX-2a, FLASH-460 and FLASH-810 respectively,
to convert antenna temperatures into main-beam temperatures. A detailed
description of APEX and its performance is given by
\citet{2006A&A...454L..13G}.

\begin{table}
\centering
\caption{Summary of the parameters of the APEX line observations.}\label{obs}
\begin{tabular}{lcccr}
\hline
\hline
\multicolumn{1}{c}{Line} &\multicolumn{1}{c}{Tuning Frequency}&\multicolumn{1}{c}{Beam}&\multicolumn{1}{c}{$\Delta v$}&\multicolumn{1}{c}{$T_{\rm sys}$}\\
\multicolumn{1}{c}{} &\multicolumn{1}{c}{(MHz)}&\multicolumn{1}{c}{($\arcsec$)}&\multicolumn{1}{c}{(km~s$^{-1}$)}&\multicolumn{1}{c}{(K)}\\
\hline

HCO$^+$$^{\mathrm{a}}$~(4-3)&356734.3&18&0.05&270\\
CH$_3$OH ($6_K-5_K$)&289125.0&22&0.06&293\\
CH$_3$CN ($16_K-15_K$)&294450.0&21&0.06&330\\
CH$_3$OH ($7_K-6_K$)&337800.0&19&0.05&160\\
\hline

\end{tabular}
\begin{list}{}{}
\item[$^{\mathrm{a}}$] with CO (3-2)
 in the lower side band.

\end{list}
\end{table}
\begin{table}
\centering
\caption{Summary of the parameters of the APEX continuum observations.}\label{obs-con}
\begin{tabular}{ccr}
\hline
\hline
\multicolumn{1}{c}{Tuning Frequency}&\multicolumn{1}{c}{Beam}&\multicolumn{1}{c}{$T_{\rm sys}$}\\
\multicolumn{1}{c}{(MHz)}&\multicolumn{1}{c}{($\arcsec$)}&\multicolumn{1}{c}{(K)}\\
\hline

344770.0&18&150\\
461040.8&14&590\\
806651.8&8&2171\\

\hline

\end{tabular}
\end{table}
\section{Observational results}\label{obs-results}

\subsection{HCO$^+$ and CO maps}
\begin{figure*}
\centering
\subfigure[]{\includegraphics[bb=205 15 551 694,clip,angle=-90,width=9cm]{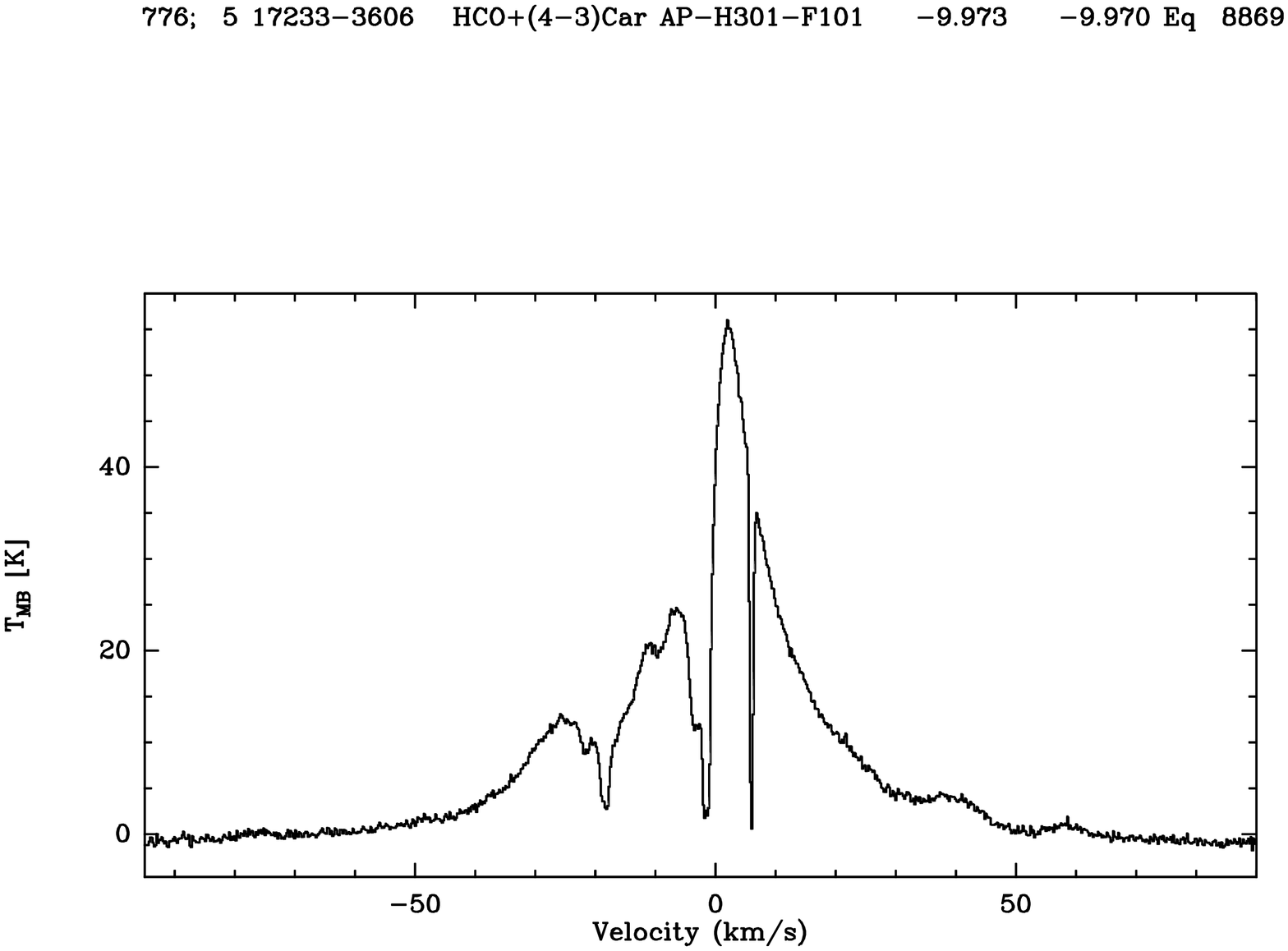}\label{co}}
\subfigure[]{\includegraphics[bb=205 15 551 694,clip,angle=-90,width=9cm]{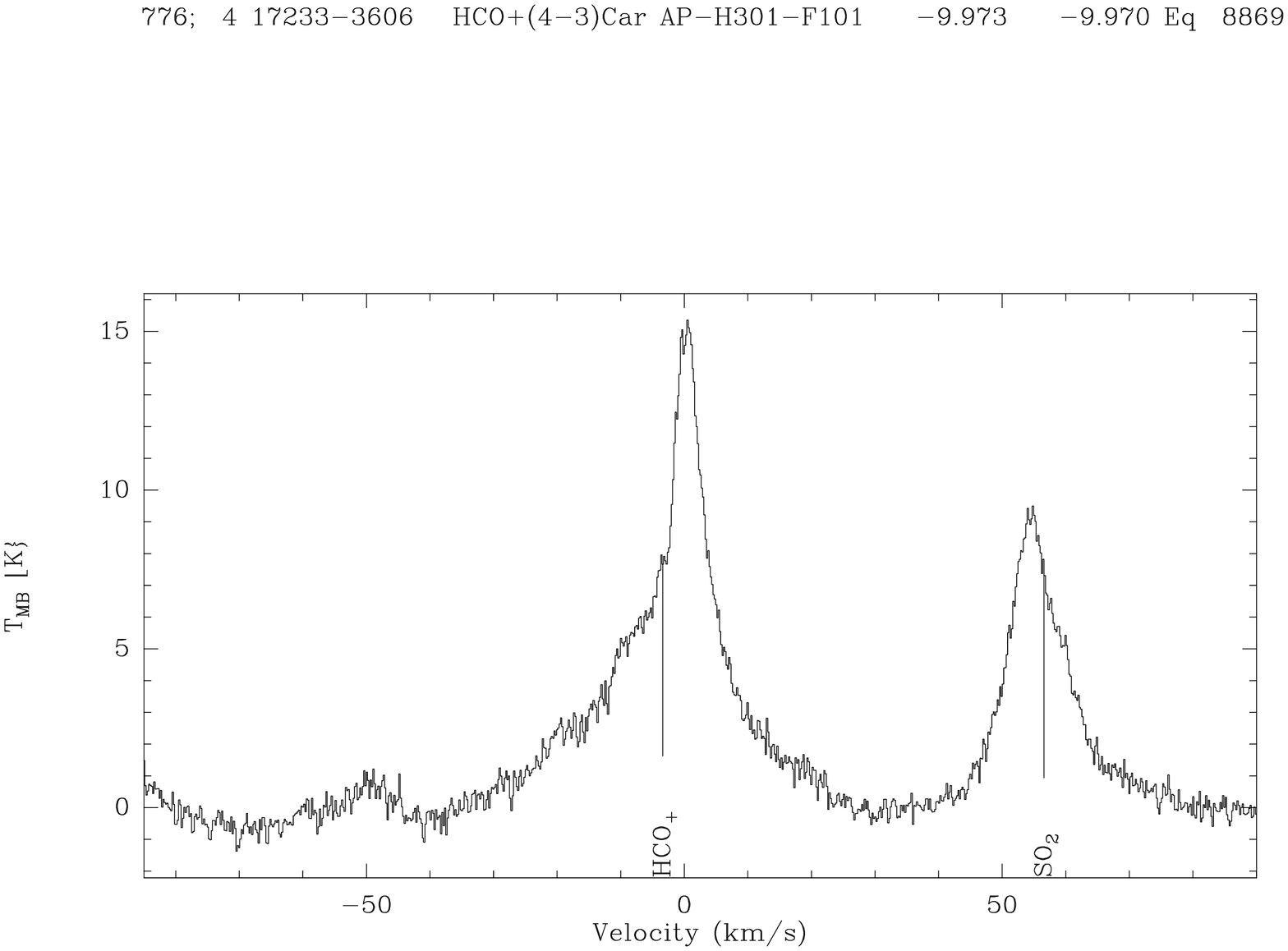}\label{hcop}}

\caption{CO (3-2) {\bf(a)} and HCO$^+$ (4-3) {\bf(b)} spectra towards the peak position of the HCO$^+$ (4-3) line
(marked by a triangle in Fig.~\ref{map}). }\label{co-hcop}
\end{figure*}

Figure~\ref{co} shows the spectrum of the CO (3-2) transition toward
the peak position of the HCO$^+$ (4-3) emission. The line, while at
the edge of the observed band, is only marginally affected by the
aliasing out-of-band problem discussed in section \S~\ref{par-obs},
since the 50~MHz at the edge of the FFTS unit starts at a velocity of
$-51$~km~s$^{-1}$ where the emission in the blue-shifted wing is
negligible.  The line has non-Gaussian blue- and red-shifted wings
with respect to the systemic ambient cloud velocity
($-3.4$~km~s$^{-1}$), which extend to a maximum blue- and red-shifted
velocity of $\sim \pm $50~km~s$^{-1}$ respectively.  In addition, two
narrower lines are detected at  $\sim 40$ and $58$~km~s$^{-1}$, which
could be high velocity bullets associated with the outflow, but
could also be due to unidentified molecular species and be
associated with the central source. The rest frequencies of these
features are 345.75 and 345.72~GHz respectively, if they come from the lower side
band of the receiver,  356.25 and 356.28~GHz if they are from the upper
side band.

\begin{figure*}
\centering
\subfigure[]{\includegraphics[angle=-90,width=7cm]{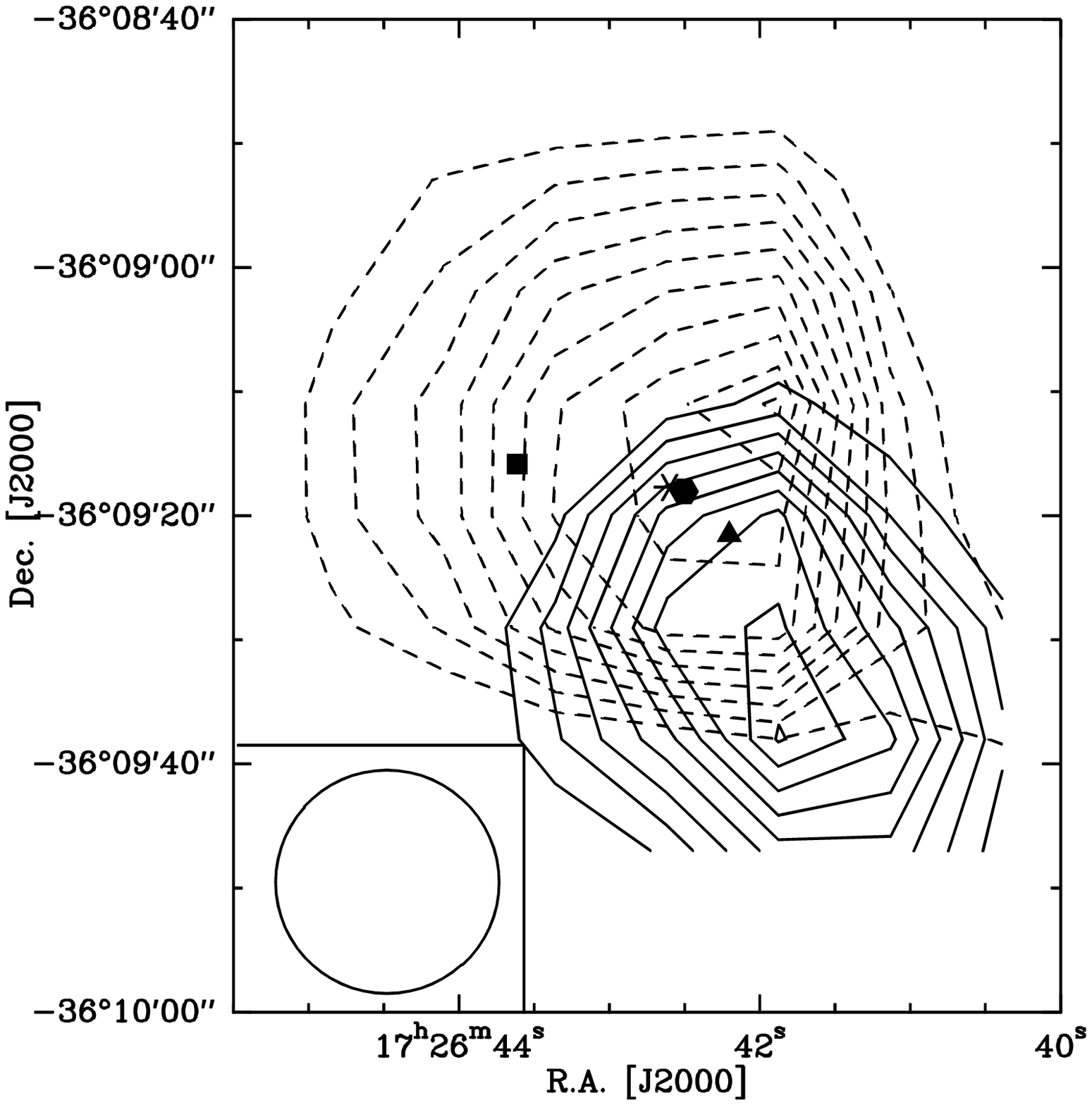}\label{co-flow}}
\subfigure[]{\includegraphics[angle=-90,width=7cm]{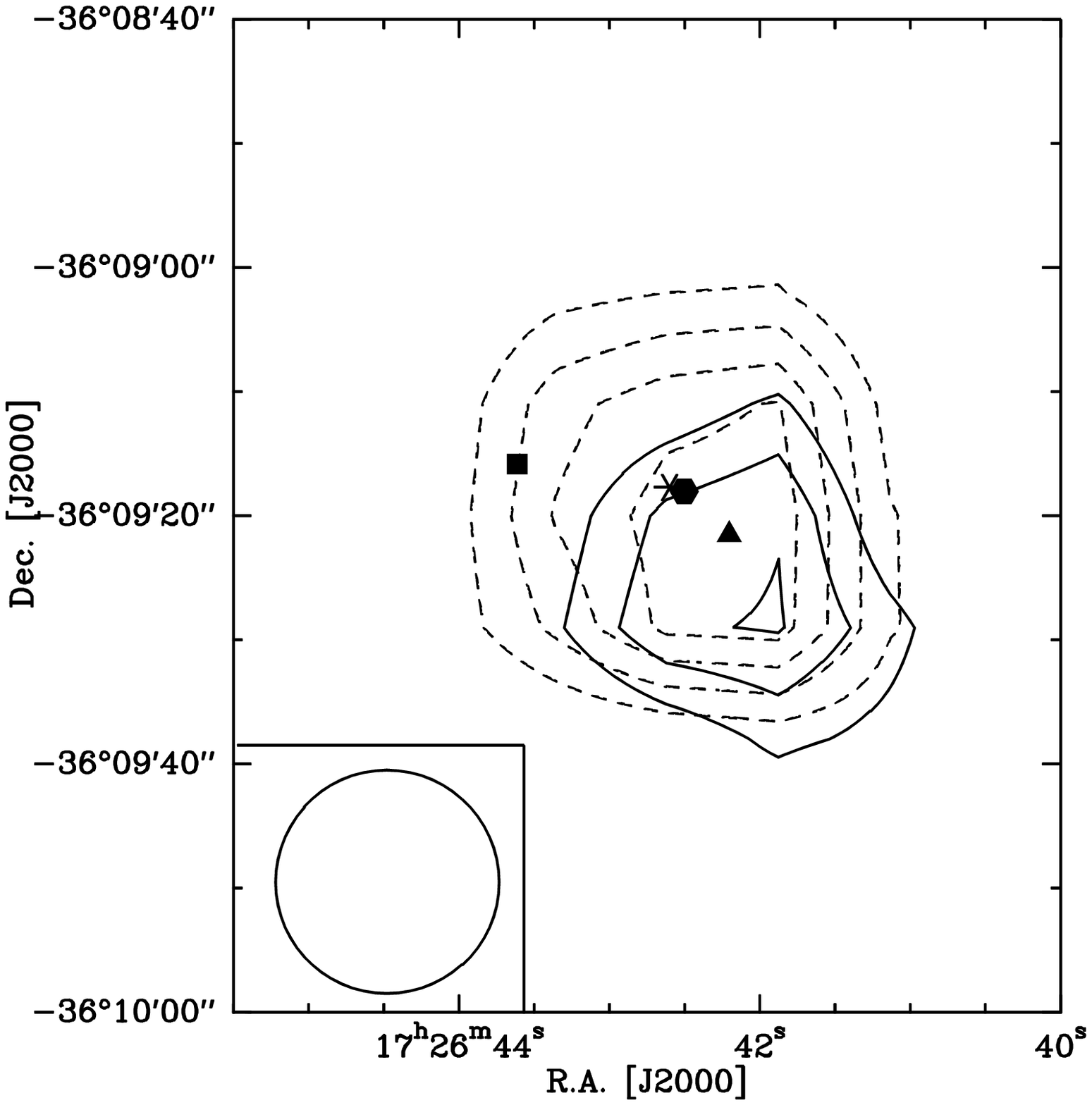}\label{hco+-flow}}

\caption{{\bf a):} Map of the integrated intensity of the CO (3-2)
transition (blue-shifted emission $v=[-50,-10]$~km~s$^{-1}$;
red-shifted emission $v=[9,50]$~km~s$^{-1}$; levels from
70~K~km~s$^{-1}$ in steps of 35, equivalent to $\sim 5~\sigma$). {\bf
b):} Map of the integrated intensity of the HCO$^+$ (4-3) transition
(blue-shifted emission $v=[-35,-8]$~km~s$^{-1}$; red-shifted emission
$v=[5,24]$~km~s$^{-1}$; levels from 20~K~km~s$^{-1}$ in steps of 10,
equivalent to $\sim 5~\sigma$). In both maps dashed contours are used
for the blue-shifted emission, solid contours for the red-shifted
emission. The triangle marks the peak position of the HCO$^+$ emission, the square 
the compact H{\sc II} region, the hexagon the centroid of the multiple radio continuum  source, and the asterisk the position of the CH$_3$OH maser spot \citep[][]{1998MNRAS.301..640W,zapata}.
The beam of the APEX telescope at the
observing frequency is shown in the left bottom
corner.}\label{map}
\end{figure*}

Broad non-Gaussian wings are also seen in the HCO$^+$
transition ($\rm{v_{blue}^{max}}=-38$, $\rm{v_{red}^{max}}=26$~
km~s$^{-1}$). However, the profiles of the two transitions differ
significantly from each other, with the CO line showing deep
absorption features probably due to self-absorption as well as to
absorption from other unrelated foreground clouds.
Figures~\ref{co-flow} and \ref{hco+-flow} show a map of the integrated
intensity of the two transitions in the line wings.

\subsection{Molecular observations}\label{spectrum}

To
investigate the physics of the molecular core in IRAS\,17233, we
performed long integration observations toward the peak position of
the HCO$^+$ emission  in several CH$_3$OH transitions and in the
CH$_3$CN ($16_K-15_K$) band, since both molecules can be used as
probes of the physics of the gas \citep{1993A&A...276..489O,2004A&A...422..573L}.
The flexibility of the FFTS at APEX
allows the simultaneous observation of 1.8~GHz bandwidth, with a
velocity resolution down to $\sim$~0.05~km~s$^{-1}$ at $\sim$1~mm. For the analysis of
the data, we smoothed all spectra to a resolution of 1~km~s$^{-1}$.
In this way, we observed transitions from the first torsionally excited levels of methanol
together with transitions from its ground state;  lines from CH$_3^{13}$CN  and $^{13}$CH$_3$OH also fall in the observed bands.
We could  detect the
CH$_3$CN ($16_K-15_K$) band up to $k=7$. The identification of higher $k$ transitions in the observed band
is rendered more difficult by their overlap with other lines. Having access to high
excitation lines, like the $\rm{v_t}=1$~CH$_3$OH transitions, and to
low opacity lines (e.g., the CH$_3^{13}$CN and the CH$_3^{13}$OH bands) guarantees a better
determination of the physical parameters of the emitting gas (see \S~\ref{core}).

\begin{figure*}\centering
\includegraphics[angle=-90,bb=169 15 551 682,clip,width=12cm]{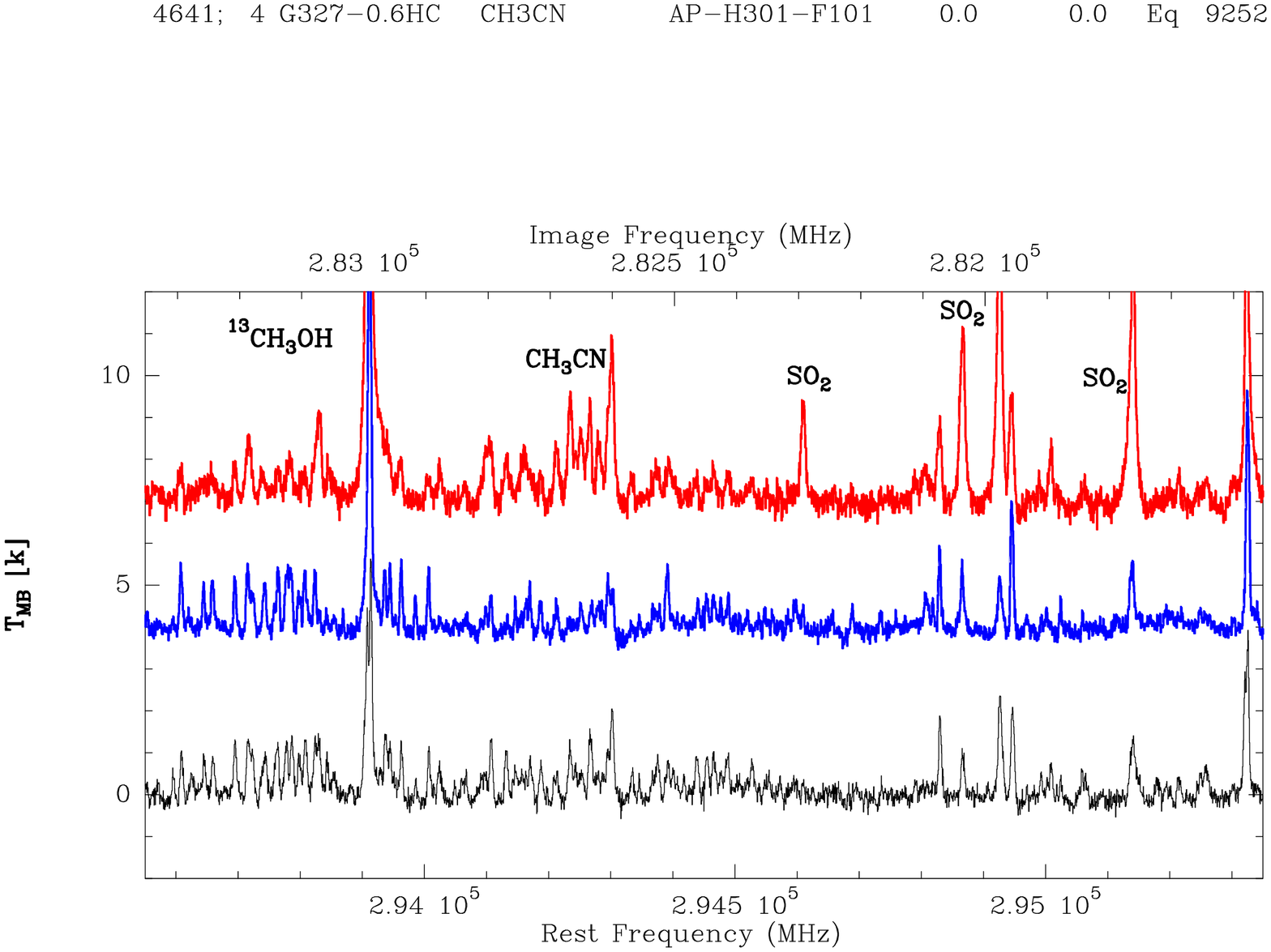}
\caption{Spectrum at 294.5~GHz of G327.3-0.6 (bottom), NGC6334(I)
(centre) and IRAS\,17233 (top). The features labelled in the spectra
are the ones discussed in section \S~\ref{comparison}. The strong
feature at the right end of the bandwidth is a mirror from the lower
side band of the H$_2$CO ($4_{1,4}-3_{1,3}$) transition at
281.5~GHz.}\label{hmc}
\end{figure*}

The molecular spectrum of IRAS\,17233 shows strong emission from
several CH$_3$OH and CH$_3$CN lines, typical of hot cores; moreover,
several other transitions from complex molecules are detected.
For comparison, we observed the southern hot cores in NGC6334(I) and
G327.3-0.6 \citep[e.g.,][]{2006A&A...454L..41S} in the same setup of
IRAS\,17233 in the CH$_3$OH ($6_K-5_K$) and CH$_3$CN ($16_K-15_K$)
bands. 
 The spectra are in general comparable,
although the linewidths of IRAS\,17233 ($\Delta v\sim
6-9$~km~s$^{-1}$) are larger than in the other two hot cores ($\Delta v\sim 3-5$~km~s$^{-1}$).
Several species  in IRAS\,17233  (CS, H$_2$CO, SO, SO$_2$, HCO$^+$, CO) show non-Gaussian
wings probably associated with the outflow.
Figure~\ref{hmc} presents one of the observed spectral window towards the three sources.

\section{Physical parameters}

 The submm continuum flux of IRAS\,17233 is strong enough to do
continuum cross-scans on the source. These observations found the peak
emission at ($-6'',-2''$) from the position used as reference,
and not coincident with the peak of the HCO$^+$ (4-3) emission (see
Fig.~\ref{map}) which we used for the spectroscopic observations of the hot core.   Fit results from averaged cross-scans at the
different frequencies are given in Table~\ref{cont}.

The peak position of the HCO$^+$ emission was derived by integrating
the line intensity over a range of velocities close to the systemic
velocity, and contamination from the molecular outflow could affect
this estimate.  The peak of the submm-continuum emission is very close
($-3''.7, 0''.2$) to the position of the methanol maser spots detected
by \citet{1998MNRAS.301..640W}, and a weak radio continuum source
\citep[named 2b by][]{zapata}.  With the expected pointing uncertainties the three positions are practically indistinguishable.
 In the following discussion, we will use the
position of the the class~II methanol maser spots
($\alpha_{2000}=17^h26^m42.6^s,~\delta_{2000}= -36^\circ09^\prime
17.7^{\prime\prime}$) as reference position for the hot core in
IRAS\,17233, since class~II methanol masers are found only in the
vicinity of massive protostars.  We also assume that the hot core is
the powering source of the molecular outflow and that it is associated
with the UCH{\sc II} region detected at cm wavelengths
\citep{1993AJ....105.1495H}. Furthermore, in the following discussion
we will assume that the total luminosity of the region is driven by
the same source (harbouring the hot core and the UCH{\sc II}).
 The spectral type of the UCH{\sc II} derived from its 
radio continuum emission is B1 \citep{zapata}, while that of 
the H{\sc II} region B0.5. However,
the 850~$\mu$m dust continuum map, taken
during the ASTROGAL survey of the Galactic Plane with the APEX
telescope (Schuller, priv. comm.),  is peaked on the hot core and
not on the H{\sc II} region, suggesting that the IR luminosity of the source is indeed dominated by the
UCH{\sc II} region.

Observations at shorter wavelengths were performed by several authors:
\citet{1982ApJ...259..657F} and \citet{1999MNRAS.309..905W} detected emission in the direction of the
UCH{\sc II} region in the near infrared $K$ and $L$ bands ($2.2\mu$m and $3.3\mu$m respectively), while the H{\sc II} region is detected
in the IR $K$ and $L$ bands and at 10~$\mu$m \citep{1982ApJ...259..657F,1999MNRAS.309..905W,2000ApJS..130..437D}.
The area is also covered by the 
GLIMPSE \citep{2003PASP..115..953B}
survey from the Spitzer Space Telescope, whose point source catalogue
has an astrometric accuracy of 0.3$''$. 
No emission is
detected in the 3.6, 4.5 or 8.0~$\mu$m bands towards the hot core, but all
bands are dominated by strong emission from a source to the north-east
of the hot core, not associated with any cm continuum emission.
Emission
in the 3.6 and 4.5~$\mu$m bands is also found close
to the hot core at ($-4.8''$, 0.2$'''$).  The MIPSGAL survey also
covers this area \citep{2005AAS...207.6333C}, but both bands (24
 and 70$\mu$m) are saturated. However, by comparing the emission at
24$\mu$m (and 70) with that at 8$\mu$m, the peak of the intensity in the MIPS bands
seems to be shifted towards the hot core position as at mm wavelengths.

\subsection{Continuum}

  The mass of the core can be derived from the measured flux at submm
wavelengths, since in this part of the spectrum the continuum emission
is dominated by thermal dust emission. However, the measurements at 461 and 806.7~GHz
should be regarded as upper limits to the continuum emission since they are likely to
be contaminated by emission from the CO(4-3) and (7-6) lines respectively.
Assuming a distance of 1~kpc
for the source, and a temperature of 70~K, the flux at 347.7~GHz
converts into a mass for the core of 107~M$_\odot$ for a dust
emissivity of 1.8~cm$^2$g$^{-1}$ \citep{1994A&A...291..943O}, or 130~M$_\odot$ for a
standard \citet{1983QJRAS..24..267H} opacity with $\beta=1.5$. The value
obtained by \citet{2004A&A...426...97F} was 210~M$_\odot$, for
$T_d=45$~K and d=800~pc.

\begin{figure}
\centering
\includegraphics[angle=-90,width=7cm]{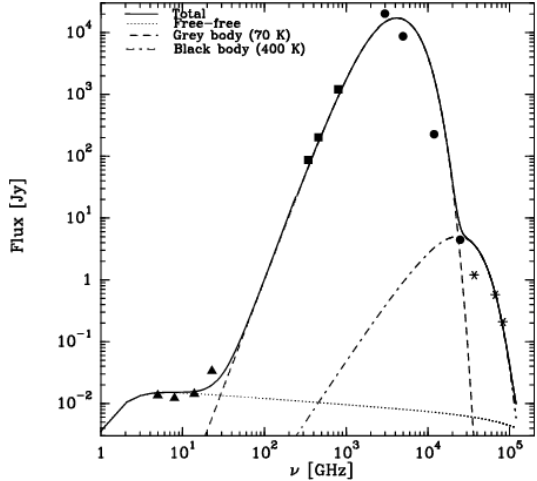}

\caption{Spectral energy distribution of the hot core in IRAS\,17233.
Observational fluxes from the VLA, APEX, IRAS and IRAC are marked
as triangles, squares, circles and asterisks respectively. For
illustration, the flux density curve of an ultra-compact H{\sc II}
region, of a 70~K grey body, and of a 400~K black body are also
given. The solid black line represents the combined spectrum of the three 
components.   }\label{sed}
\end{figure}

 In
Fig.~\ref{sed}, we show the spectral energy distribution of
IRAS\,17233 from 5~GHz to the near IR.
Given our resolution, several of the IRAC
sources in the region could coincide with the hot core giving rise to
the submm emission or none at all. Since we assume that the hot core
coincides with the CH$_3$OH maser, where no IRAC emission is detected,
 we actually believe the latter to be
the case. However, we included the fluxes in the IRAC bands in Fig.~\ref{sed}, integrated 
over the APEX beam.
The cm fluxes are from the data presented by \citet{zapata};
they correspond to the sum of fluxes of their sources VLA2a--d, restored with a beam of $\le 1''$.
An important difference between the VLA, APEX and IRAS
observations is the spatial resolution, which is  high in the cm
region  and coarse at the shorter wavelengths. From
the source sizes derived from the analysis of the submm continuum (see
Table~\ref{cont}), the emission detected with APEX is not confined to
a compact source of a few arcsec or less, as the VLA
sources, but it is rather extended, and probably contaminated by emission
from components other  than the hot core, e.g., a surrounding envelope.
Therefore, one should regard the flux measures from APEX as
upper limits to the fluxes of the compact hot core associated with the
cm free-free emission.

\begin{table}
\centering
\caption{Summary of the APEX continuum observations.}\label{cont}
\begin{tabular}{crccr}
\hline
\hline
\multicolumn{1}{c}{Frequency}&\multicolumn{1}{c}{T$_{\rm{MB}}$}&\multicolumn{1}{c}{Obs. size}&\multicolumn{1}{c}{Dec. size}&\multicolumn{1}{c}{Flux density}\\
\multicolumn{1}{c}{(GHz)}&\multicolumn{1}{c}{(K)}&\multicolumn{1}{c}{($\arcsec$)}&\multicolumn{1}{c}{($\arcsec$)}&\multicolumn{1}{c}{(Jy)}\\
\hline

347.7&1.8&22.1&12.7&88\\
461.0&3.6&17.9&12.0&204\\
806.7&10.7&14.6&12.4&1214\\
\hline
\end{tabular}
\end{table}

\subsection{Outflow}

\begin{figure}
\centering
\includegraphics[bb= 114 139 527 613, clip,angle=-90,width=7cm]{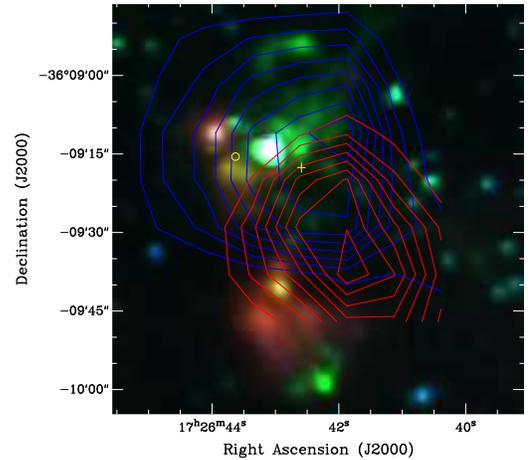}
\caption{Spitzer image of the IRAS\,17233$-$3606 region; the colour scheme is
the following: 3.6$\mu$m (blue), 4.5~$\mu$m (green) and 8$\mu$m (red). The blue and red contours represent the integrated intensity
of the CO(3-2) transition in the blue- and red-shifted wings
respectively. Contours level are as in Fig.~\ref{co-flow}. The yellow
cross shows the position of the CH$_3$OH maser spot, the square
the compact H{\sc II} region \citep{1998MNRAS.301..640W}.}\label{glimpse}
\end{figure}

 Figure~\ref{co-flow} shows the map of the molecular outflow originating from IRAS\,17233 traced by the CO(3-2) line; plotted
are the intensities integrated over the velocity ranges covered by velocities
lower and higher than the systematic velocity, i.e, red- and blue-shifted emission, respectively.
 Although the
spectrum at the central position (see Fig.~\ref{co}) has both red and blue-shifted
emission at high velocities, implying that the inclination angle of
the flow with respect to the line of sight is small, the integrated
intensity of the CO and HCO$^+$ lines in the wings shows a bipolar distribution: these two
properties suggest a medium value for the inclination angle of the outflow.
The flow has a north-south orientation, with
the blue-shifted emission  towards the south, and the
red-shifted emission to the north. Strong emission at
4.5~$\mu$m, which is often interpreted as due to outflows \citep[e.g.,
][]{2004ApJS..154..352N}, is detected along the direction of the
molecular outflow (Fig.~\ref{glimpse}).

From the CO map it is clear
that the lobes are barely resolved by our observations, and that
higher resolution is needed to determine parameters such as the
collimation factor and the opening and inclination angles of the outflow.
The position angle of the flow, computed from the peak positions of the CO integrated emission
in the two lobes, is
close to $0^\circ$.

To compute the physical parameters of the outflowing gas we followed
the standard formalism described by \citet{1990ApJ...348..530C}.  The
momentum $(p)$, the energy $(E)$, and the characteristic time scale
$(t)$ can be derived assuming that the outflow is a constant velocity
flow with true velocity equal to the maximum velocity observed. For
the estimate of the mass in the lobes, we used a constant value for
the integrated intensity of the CO line equal to its mean value computed on each
lobe starting from  the 5$\sigma$ level, and assumed that this is indeed representative of the flow
region.  To avoid contamination from the core emission, the H$_2$
column density was obtained assuming a relative abundance of CO to
molecular hydrogen of 10$^{-4}$ and integrating between 9 and
50~km~s$^{-1}$ for the red-shifted gas, and -50 to -10~km~s$^{-1}$ for
the blue-shifted gas, which results in lower limits to the derived
masses. Moreover, part of the red lobe emission is not covered by our
observations.  Since similar studies on molecular outflows from
massive young stellar objects found gas in the flow at relatively high
temperatures \citep{2002A&A...387..931B,2006A&A...454L..83L}, we assume
that the excitation temperature of the outflow gas is 50~K.  Some
authors \citep[e.g.,][]{1996A&A...311..858B} apply a multiplication
factor of 3.5 for optical depth effects and a mean inclination angle of
57$^\circ$.3 in computing the outflow parameters. Since the
orientation of the outflow cannot be determined from our data, we
decided not to correct for the inclination angle of the
outflow. Moreover, observations of CO(3-2) and $^{13}$CO(3-2) towards
other massive flows show that the ratio of these two lines may change
considerably from object to object
\citep[e.g.,][]{2006A&A...454L..83L,2007A&A...463..217G} thus implying
differences in the optical depth of CO.  The values reported in
Table~\ref{outflow} should be considered as lower limits to the true
values as {\it 1)} no corrections for the optical depth or inclination
angle were applied; {\it 2)} the gas at low velocities was not
considered in the calculations.

\begin{table}
\centering
\caption{Parameters of the CO molecular outflow.}\label{outflow}
\begin{tabular}{lr}
\hline
\hline
\multicolumn{1}{c}{} &\\
$M_{\rm{tot}}$ $[M_\odot]$&0.8\\
$p[M_\odot~\rm{km~s^{-1}}]$&25.3\\
$E[10^{45}$~erg]&8.6\\
$t$ [yr]&6200\\
$\dot{M}_{\rm tot}$ [$M_\odot~\rm{yr}^{-1}$]&0.005\\
$F_{\rm m}$ [$~M_\odot~\rm{km~s^{-1}}\rm{yr^{-1}}$]&0.04\\
$L_{\rm m}$ [$L_\odot$]&48\\

\hline
\end{tabular}
\end{table}

 \citet{2002A&A...383..892B} compared the physical properties of
high-mass outflows to those of low-mass flows by studying the
correlations between the mass and  the luminosity of the core as
function of the properties of the flow. The flow in IRAS\,17233 has
physical parameters  characteristic of molecular flows driven by
high-mass young stellar objects.  In particular, the values derived
for the mass entrainment rate $\dot{M}_{tot}$ and for the mechanical
force $F$ are in agreement with a relatively low luminosity ($L\sim
10^4~L_\odot$) for the powering force of the flow.

\subsection{Molecular core}\label{core}
The analysis of the molecular spectrum of IRAS\,17233 was carried out
with the XCLASS program \citep[discussed by][]{2005ApJS..156..127C},
which uses an LTE model to produce synthetic spectra, and compares
them to the observations. The molecular data are from the CDMS
\citep{2001A&A...370L..49M} and JPL
\citep{pickett_JMolSpectrosc_60_883_1998} databases. The parameters
defining the synthetic spectrum are: source size, rotation
temperature, column density, velocity offset and the velocity
width. In our case, we did not use any offset in velocity, but
adopted the systemic velocity of the source ($v_{\rm
LSR}~-3.46$~km~s$^{-1}$) for all molecular species. Several velocity
components, which are supposed to be non-interacting (i.e. the
intensities add up linearly), can be used. The program can also
produce double side band synthetic spectra for comparison against the
observations.  This is needed to model our data, since their
frequency coverage does not permit sideband deconvolution.  The
galactocentric distance of the source was computed by
\citet{1996A&A...305..960C}, and be found equal to 7.8~kpc.
Following
\citet{1994ARA&A..32..191W}, the isotopic ratio $^{12}\rm{C}/^{13}\rm{C}$ at
this distance is equal to 66, very close to the standard value
of 60 used in XCLASS, and well within the errors of this estimate.

We started our analysis with the modelling of the CH$_3$CN spectrum
because several transitions are detected at once, providing good
constraints on
the physics of the gas they trace.
The corresponding CH$_3^{13}$CN band also
falls  within the bandwidth of our
observations; however, at the rest frequencies of these lines,
blending
with other molecular species is severe and it is
difficult to assess whether some of the lines are detected or not.  However,
even their non-detection would give us useful upper limits on the
column density of this species.  The best fit was found for a source
size of 2.7$''$, a temperature of 150~K and a column density of $\sim
4\times 10^{16}$~cm$^{-2}$.

 \citet{1995ApJ...449L..73W} noticed that for luminosities typical of massive young stars,
 the temperature distribution in spherical, centrally illuminated dust clouds can be well
approximated by the relation

\begin{equation}
T_d(R)=37\left(\frac{L}{L_\odot}\right)^{0.25}\left(\frac{R}{100~AU}\right)^{-0.4} [\rm {K}]\label{wilner},
\end{equation}

which relates the temperature of the
dust ($T_d$) at a given distance ($R$) to the
luminosity ($L$) of the inner source.
Using this equation, and assuming local thermal equilibrium between
dust and gas, we can independently derive the typical angular size for the hot core in IRAS\,17233.
\citet{1993AJ....105.1495H} estimated an IR luminosity for the source of $1.5\times
10^5$~L$_\odot$, using a distance of 2.2~kpc. This estimate decreases
to $\sim 2.5\times 10^4$~L$_\odot$ when adopting a distance of 1~kpc \citep{1998AJ....116.1897M},
a value similar to that of \citet{ 2004A&A...426...97F}, who derived a
bolometric luminosity of 1.4$\times 10^4$~L$_\odot$ with a distance of
0.8~kpc.

Assuming a luminosity of $\sim 1.4\times 10^4$~L$_\odot$,
the diameter of the region  at 150~K is
2700~AU for IRAS\,17233  (see discussion in \S~\ref{comparison}),
which corresponds to 2.5$''$ at a distance of 1~kpc, in agreement with the value
derived through the analysis of the molecular spectrum.

\begin{figure*}
\centering
\includegraphics[angle=-90,width=13cm]{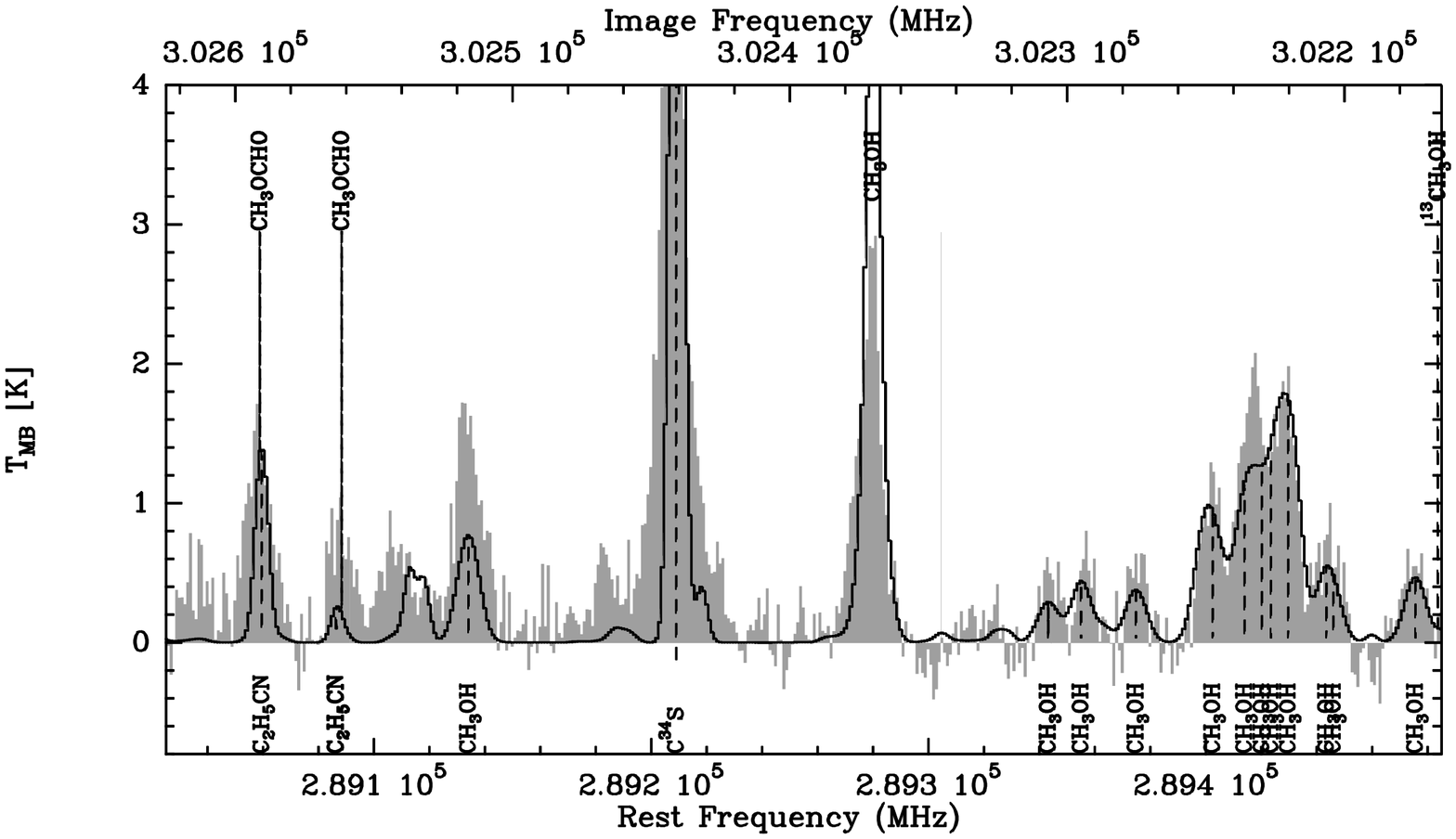}
\includegraphics[angle=-90,width=13cm]{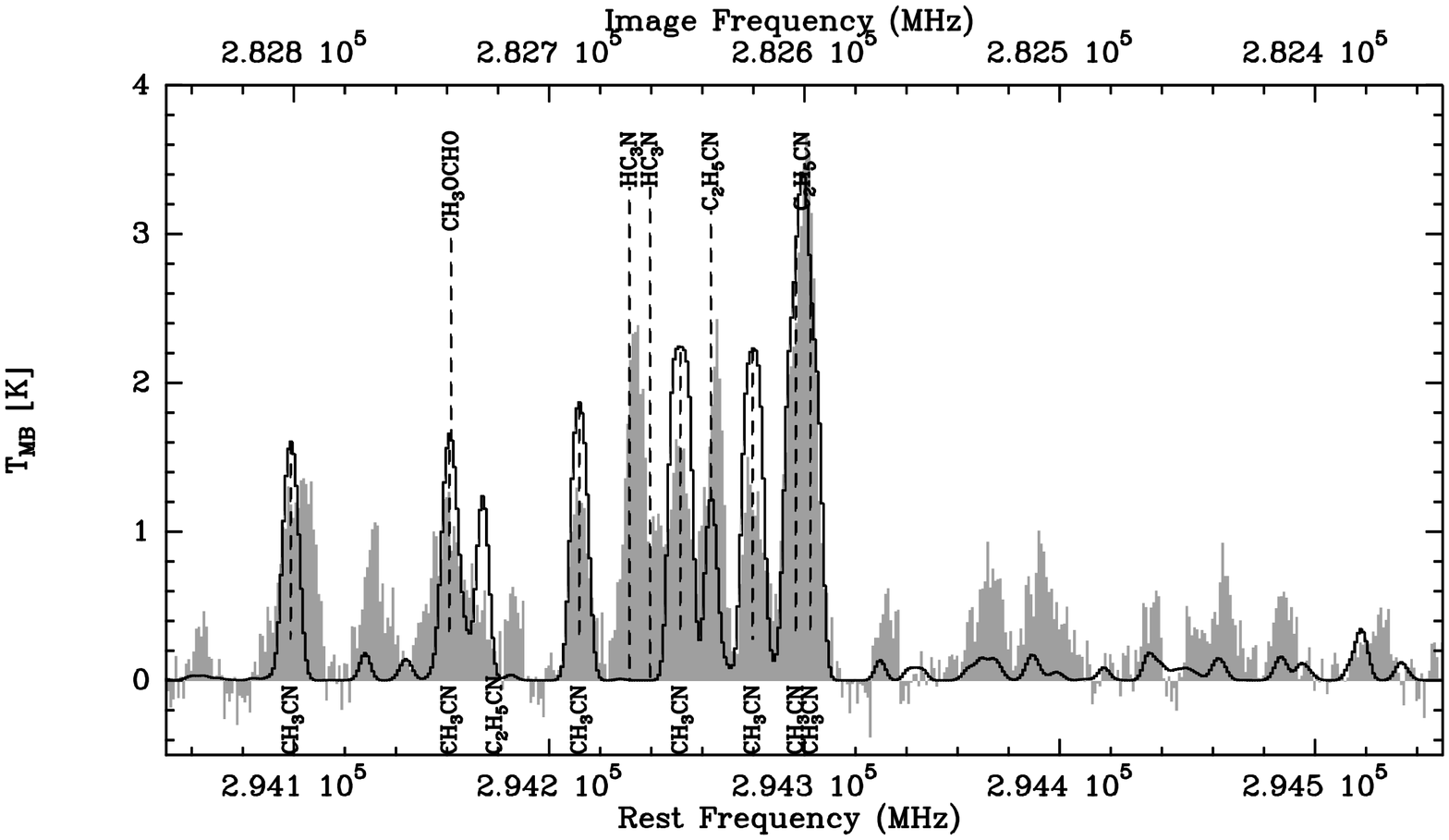}
\caption{Molecular emission from IRAS\,17233$-$3606. The solid line
represents the synthetic spectrum obtained for the best fit
solution.}\label{ch3oh_ch3cn}
\end{figure*}

\begin{figure*}
\centering
\includegraphics[angle=-90,width=13cm]{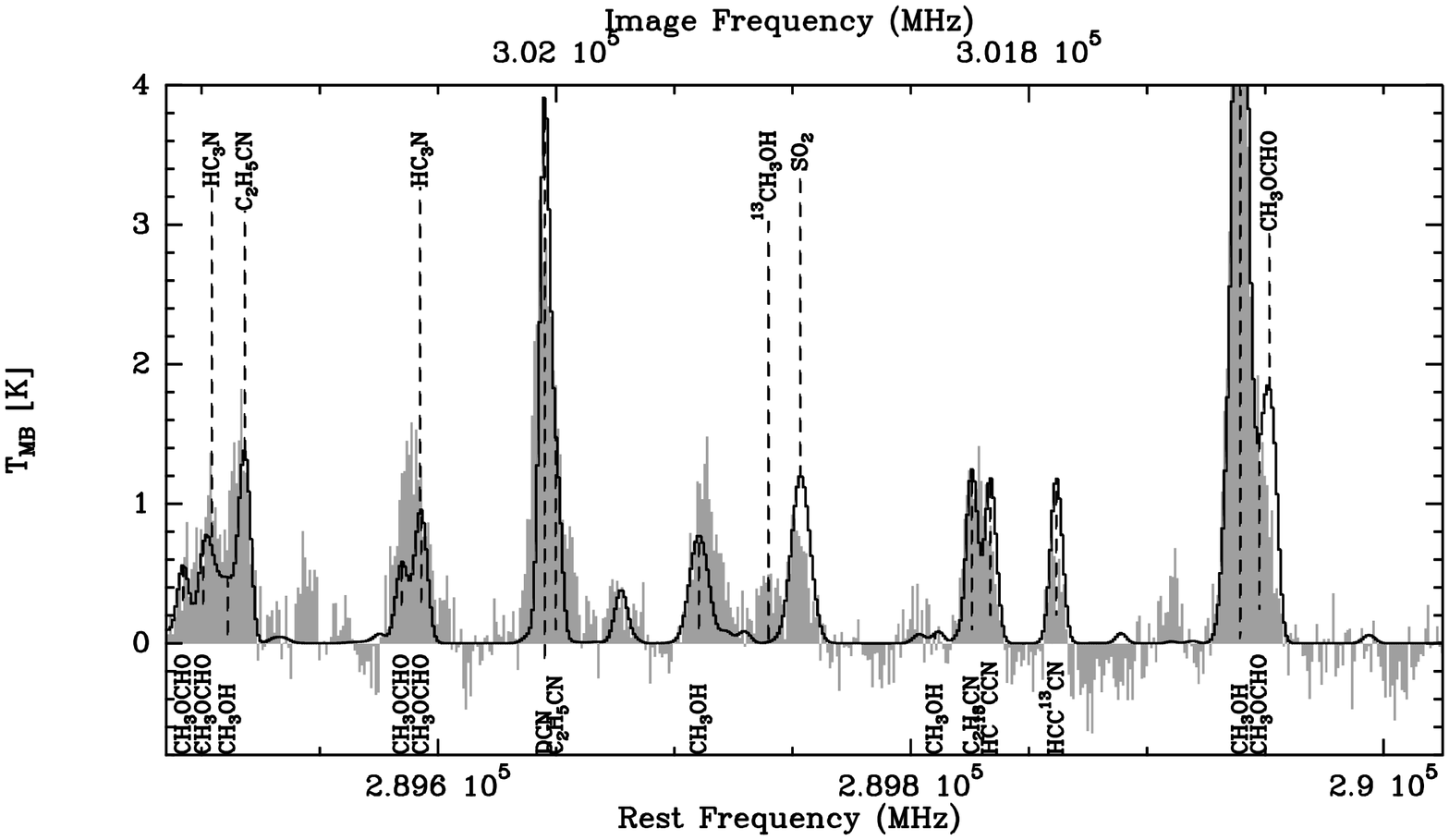}
\includegraphics[angle=-90,width=13cm]{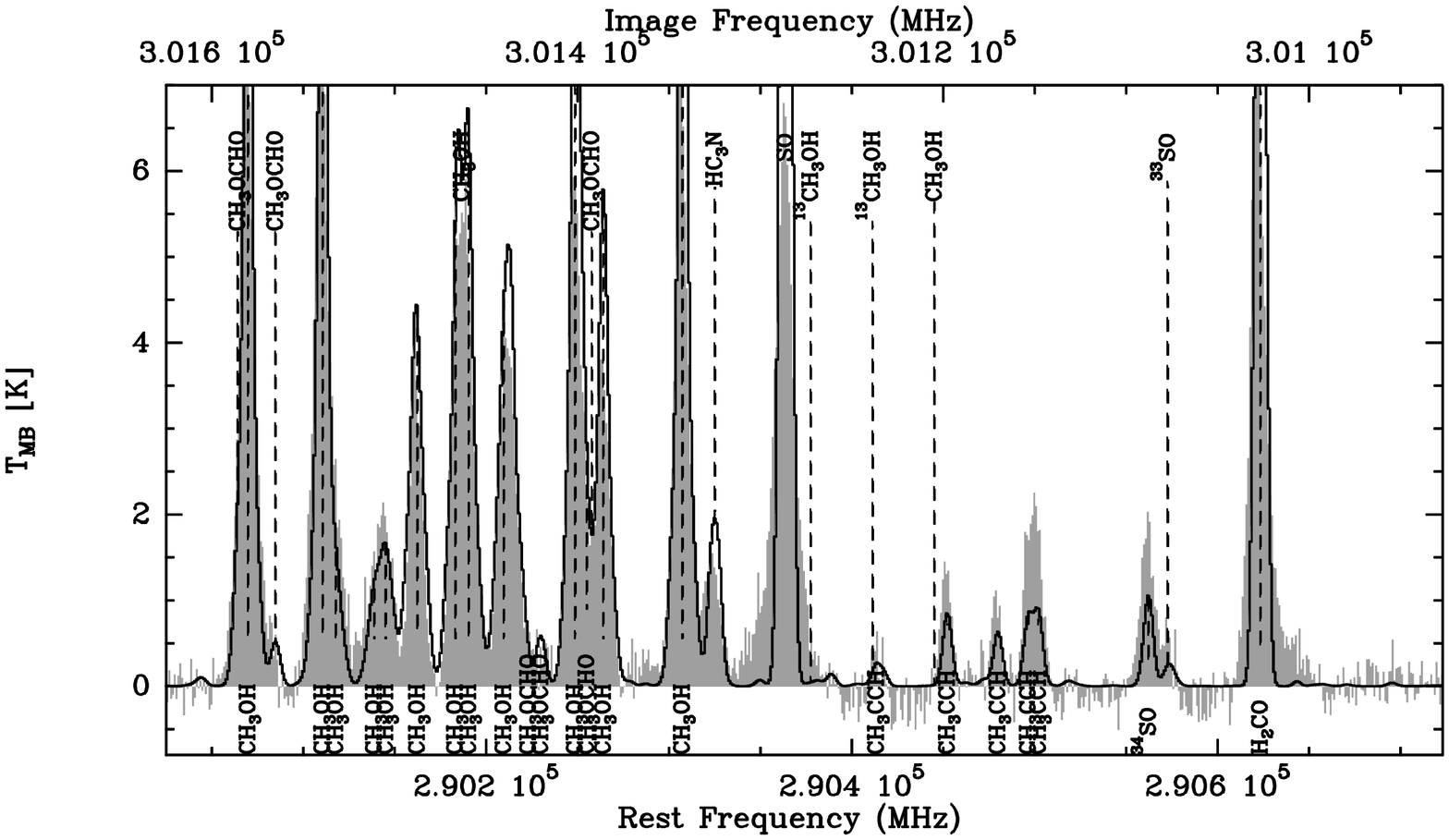}
\includegraphics[angle=-90,width=13cm]{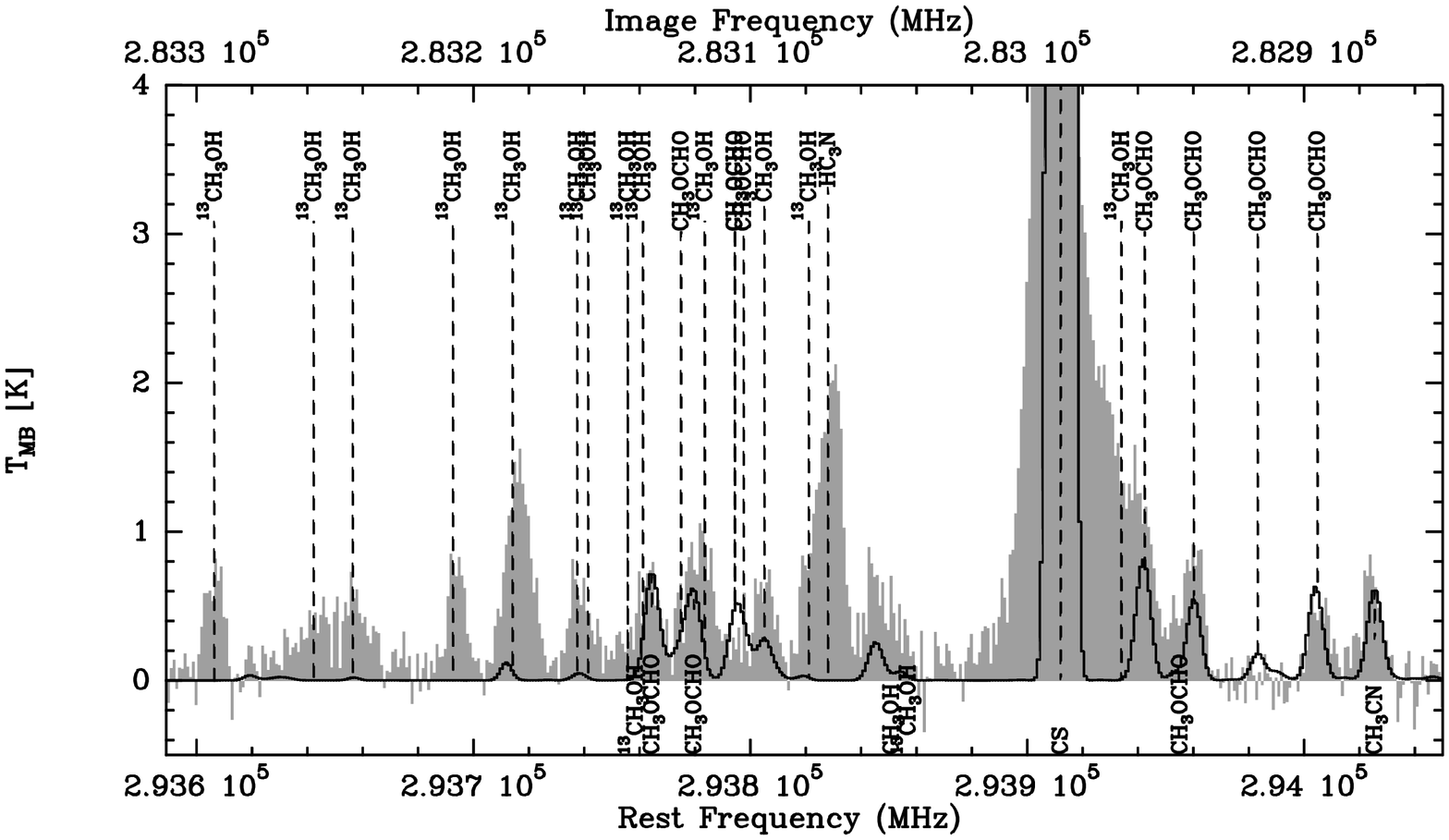}

\caption{Molecular emission from IRAS\,17233$-$3606. The solid line
represents the synthetic spectrum obtained for the best fit
solution.}\label{otherspectra}
\end{figure*}

For the best fit values, most of the CH$_3$CN
lines are optically thick; this implies a degeneracy in the model
between the source size and the temperature.
To derive the $3\sigma$
confidence levels in the parameters, we fixed the source size to the
best value derived in our analysis, and performed a $\chi^2$ analysis
for the other two parameters  in the spectral window between 294.0  and 294.3~GHz, which contains
the  CH$_3$CN and CH$_3^{13}$CN $(16_K-15_K)$ lines. We found that the 3~$\sigma$ confidence level for
 the kinetic temperature of
the gas and the column density ranges between 114 and 290~K, and
2$\times 10^{16}$ and 6$\times 10^{16}$~cm$^{-2}$, respectively.  We performed the
same analysis on the torsionally excited lines of methanol.  Although
excitation by collisions is not an efficient way to populate these
levels, which are pumped through the infrared radiation field \citep{2007A&A...466..215L}, their
populations are easily thermalised at the temperature of the dust by
the dust radiation field. Since it can be assumed that dust and gas are in
thermal equilibrium \citep{1984A&A...130....5K} at the high density of the inner
region around massive protostars, we believe that the
LTE approach, defined by one temperature for the dust and for the gas,
is reasonable also for the $v_t=1$ lines of methanol.  The
minimum of the $\chi^2$ distribution is found for different values
than the ones delivered by the CH$_3$CN analysis (T$\sim 210$~K,
s$\sim1.7''$, N$\sim 9\times 10^{18}$~cm$^{-2}$). However, this model
offers a good fit for the higher energy levels in the band, while the
other lines have high optical depths and their profiles do not match
the observations. Therefore, in Table~\ref{results} we list the column
density obtained for a source size and a temperature equal to that
obtained from the CH$_3$CN analysis, which still fall in the
$3~\sigma$ confidence level of the $\chi^2$ distribution of CH$_3$OH.
In Fig.~\ref{ch3oh_ch3cn}, we present an example of the quality of the fit for CH$_3$OH  $v_t=1$
and CH$_3$CN lines. The other spectral windows, and the synthetic spectrum corresponding
to the best fit model, are shown in Fig.~\ref{otherspectra} in the Online Material section.

The parameters used in the model for the different molecular species are shown in
 Table~\ref{results}.  The emission of several molecular species is
 compatible with that of a compact source with the temperature and
 source size derived from acetonitrile.  These are complex molecules,
 whose abundances are enhanced by thermal evaporation of grain mantles,
 or whose high excitation lines are efficiently excited only by
 infrared pumping.  In the remaining cases,
 the source
 was assumed to be extended, and the column density
 given in Table~\ref{results} corresponds to the beam averaged column
 density. This approach is justified for optically thin emission,
 where the source size and the column density are degenerate
 parameters of the model. However, it should be noticed that this
 approach was used mostly for molecules where only a limited number of
 transitions were available.

\begin{table}
\centering
\caption{Molecular parameters for IRAS\,17233.}\label{results}
\begin{tabular}{lrrcc}
\hline
\hline
\multicolumn{1}{c}{Species} &\multicolumn{1}{c}{Size}&\multicolumn{1}{c}{$T_{ex}$}&\multicolumn{1}{c}{$N$}&\multicolumn{1}{c}{Comment}\\
\multicolumn{1}{c}{} &\multicolumn{1}{c}{$(\arcsec)$}&\multicolumn{1}{c}{(K)}&\multicolumn{1}{c}{(cm$^{-2}$)}\\

\hline
C$_2$H$_5$CN    &  2.7$^a$ &     150$^a$ &  5$(16)$&$^{\mathrm{b}}$\\
CH$_3$CCH       &  ext.$^a$ &     100 &  2$(15)$&$^{\mathrm{c}}$\\
CH$_3$CN        &  2.7 &     150 &  4$(16)$\\
CH$_3$OCHO-a    &  2.7$^a$ &     150$^a$ &  2$(17)$&$^{\mathrm{b}}$\\
CH$_3$OCHO-e    &  2.7$^a$ &     150$^a$ &  2$(17)$&$^{\mathrm{b}}$\\
CH$_3$OH        &  2.7$^a$  &     150$^a$ &  2$(18)$&$^{\mathrm{d}}$\\
                &  ext.$^a$  &      40 &  8$(15)$&\\
CO              &  ext.$^a$  &      30 &  5$(19)$&$^{\mathrm{e}}$\\
CS              &  ext.$^a$  &      30 &  5$(15)$&$^{\mathrm{c}}$ \\
DCN             &  ext.$^a$  &     150 &  4$(13)$&$^{\mathrm{c}}$ \\
H$_2$CO         &  ext.$^a$  &      80 &  2$(15)$&$^{\mathrm{c}}$ \\
H$_2$CS         &  ext.$^a$  &      70 &  9$(14)$&$^{\mathrm{c}}$ \\
HC$_3$N         &  2.7$^a$  &     150$^a$ &  3$(16)$ \\
                &  ext.$^a$   &      50 &  9$(15)$ \\
HNCO            &  2.7$^a$ &     150$^a$ &  $3(16)$&$^{\mathrm{c}}$ \\
SO              &  ext.$^a$  &      40 &  3$(15)$\\
SO$_2$          &  ext.$^a$  &      60 &  5$(15)$\\

\hline
\end{tabular}
\begin{list}{}{}
\item[$^{\mathrm{a}}$] Fixed parameter;
\item[$^{\mathrm{b}}$] Based on weak or partially blended lines only;
\item[$^{\mathrm{c}}$] Based on one or a few lines only;
\item[$^{\mathrm{d}}$] Includes vibrationally excited lines;
\item[$^{\mathrm{e}}$] Based on C$^{17}$O (3-2) only.

\end{list}
\end{table}

\subsection{Comparison with other hot cores}\label{comparison}

 Comparison between hot cores associated with sources of different
luminosities can help to understand whether their chemical composition
is a function of the mass and luminosity of the central heating
source, whether it simply reflects the evolutionary state of the
source and its initial conditions.

The assumed luminosity of   IRAS\,17233 is $\sim 1.4\times 10^4$~L$_\odot$.
For NGC6344(I) and for G327.3-0.6, attempts to derive their bolometric
luminosity were made by \citet{2000A&A...358..242S} and
\citet{2006A&A...454L..91W} respectively, who report values of
2.6$\times10^5$~L$_\odot$ for NGC6334(I), and
5--15$\times10^4$~L$_\odot$ for G327.3-0.6.
 Therefore, IRAS\,17233 is the least powerful of the
three hot cores we observed. On the other hand, it is likely to be the
closest among these sources \citep[$D_{\rm{G327.03}} \sim 2.9$~kpc,
$D_{\rm{NGC6334(I)}} \sim 1.7$~kpc,
][]{1992PhDT.......252B,1978A&A....69...51N}, and it could have the
largest beam filling factor in our observations. This would explain
its exceptionally strong molecular spectrum despite its relatively low
luminosity.

However, using Eq.~\ref{wilner},
we found that the size of the region at 150~K is equal to 2700~AU for IRAS\,17233 ($\sim 2.5''$), 15000~AU for NGC6334(I)
($\sim 8''$), and, given its luminosity
uncertainty,  between  5200 and 10\,000~AU ($\sim 1.7''$ and 3.4$''$, respectively)
for G327.3-0.6.
According to these results, the
exceptionally strong molecular spectrum of IRAS\,17233 cannot be
attributed to a beam filling factor larger than for the other sources.
This would rather suggest that the strength of the
lines from
IRAS\,17233 are due to an intrinsic
property of the source.  However, the source sizes derived through
Eq.~\ref{wilner} are affected by the uncertainties in the distance,
which at least in the case of IRAS\,17233 is high, and in the luminosity, which for all sources
is based on relatively low resolution data.

\citet{2006A&A...454L..41S} and \citet{2005ApJS..156..127C} studied
the hot cores in NGC6334(I),
G327.3-0.6 and Orion-KL with the same technique used
for IRAS\,17233, although at different frequencies. Therefore, the
results of these analyses are affected by the same systematic
errors of the method, and comparison with our results should be
straightforward.  However, the source size used in their studies is
not estimated from observations but derived from the models. Since
this parameter is degenerate with the temperature for optically thick
lines, and with the temperature and the column density for optically
thin lines, the column densities derived for the hot cores are
strongly dependent on the source size used in the modelling. To cancel
out this effect, we compared relative abundances in the different
sources. Given the different spectral windows analysed, only CH$_3$CN is observed in all hot cores.  Therefore,
we used CH$_3$CN as reference molecule to compute the abundances of
the other species.  The  species associated with the hot cores,
and in common between different sources, are CH$_3$OH and HNCO (in
IRAS\,17233, NGC6334(I), Orion-KL), C$_2$H$_5$CN (in IRAS\,17233,
Orion-KL, G327.3-0.6), and CH$_3$OCHO (in IRAS\,17233, NGC6334(I),
G327.3-0.6).  Moreover, \citet{2002A&A...390.1001S} studied the molecular
content of the hot core around the solar type protostar IRAS\,16293-2422, and derived
abundances for four of the species analysed here (CH$_3$CN, CH$_3$OH, HNCO and CH$_3$OCHO).

\begin{figure}
\centering
\includegraphics[angle=-90,width=9.5cm]{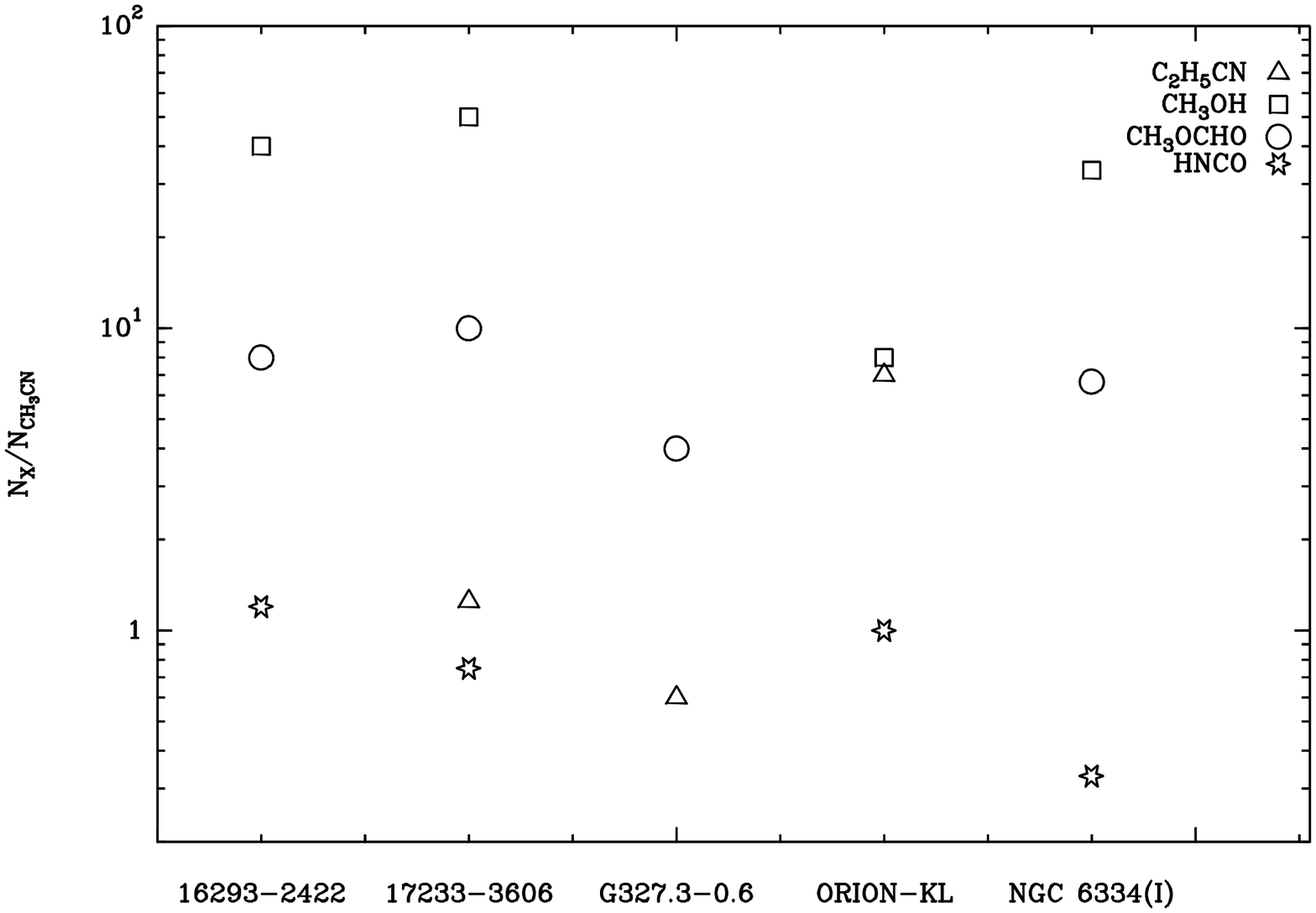}
\caption{Comparison of molecular abundances of CH$_3$OH, HNCO, CH$_3$OCHO and C$_2$H$_5$CN respect to CH$_3$CN.}\label{abu}
\end{figure}

By comparing the results for these five sources, we could not find any
trend in the abundances as function of the luminosity of the object,
and all sources show different values for each species. However, for
all sources the most abundant species is CH$_3$OH, followed in by
CH$_3$OCHO, C$_2$H$_5$CN and HNCO. For IRAS\,16293-2422, the abundance
of CH$_3$OCHO is an upper limit.  Figure~\ref{abu} presents an
overview of the molecular abundances with respect to CH$_3$CN in the five
sources.  \cite{1998A&AS..133...29H} performed a similar analysis on a
larger sample of hot cores, and found no systematic difference between
their sources. However, we did not include their results in our
analysis because for most of the molecular species they are limits
to the column densities.

We believe that such a comparison of the chemistry of hot cores
associated with heating objects of different mass and luminosity, and
including hot corinos around low mass protostars, would  indeed be
useful for a better understanding of this evolutionary stage. However,
we stress that this analysis should be performed with the same method
and on observations of the same frequency windows and linear
resolutions to cancel out all systematic errors.

 Figure~\ref{hmc} presents the molecular spectrum at $\sim 294$~GHz of the three hot cores 
IRAS\,17233,  NGC6344(I) and  G327.3-0.6. The spectra are relatively similar, although
 NGC6334(I) and G327.3-0.6 show stronger
emission than IRAS\,17233 in the $^{13}$CH$_3$OH ($6_K-5_K$)~v$_t=0,1$ band, while
the  CH$_3$CN ($16_K-15_K$) is more intense in
IRAS\,17233, at least in its lower transitions lines. However, a striking difference between the three
sources is in the intensity of the  SO$_2$ lines (and SO transitions at other frequencies), which
are by far stronger in IRAS\,17233 than in the other two cores.
The abundance of S-bearing molecules can be
enhanced by thermal evaporation of icy grain mantles due to
protostellar heating \citep[e.g., ][]{1997ApJ...481..396C}, or by
shocks \citep[e.g., ][]{1984ApJ...287..665M}.  Among other hot cores,
Orion-KL is known to be exceptionally strong in SO and SO$_2$
\citep[e.g., ][]{1997ApJS..108..301S}, which are associated with a
molecular outflow \citep{1996ApJ...469..216W}.  In the case of
IRAS\,17233, the linewidths of the SO and SO$_2$ transitions are broad
(9--12~km~s$^{-1}$), but reasonably fitted with single Gaussian
profiles, suggesting that the bulk of the emission does not arise in
the outflow.  However, for two of the SO$_2$ transitions a
two component Gaussian fit  gives a better agreement with the
observations. Both transitions have asymmetric profiles, with non-Gaussian
blue-shifted emission.
Figure~\ref{hcop} shows one of the two asymmetric SO$_2$
transitions, which is in the same frequency setup of the HCO$^+$(4-3)
observations, but comes from the other side band.
The integrated emission of this line is compact, but slightly elongated along the axis of the molecular
outflow.  However, only a map of the region in these two molecular
species at higher angular resolution would unambiguously associate
their emission with the hot core or with the outflow.

\citet{1998A&A...338..713H} and \citet{2003A&A...412..133V} studied
several massive star forming regions in different S-bearing molecules,
and found that the abundances of SO and SO$_2$ are the ones with the
largest spread from source to source.  \citet{2003A&A...412..133V}
also found that up to 50\% of the SO and SO$_2$ emission arises in
high-velocity gas. No trend in their column densities with temperature
was found. These studies agree on the fact that the abundances
of SO and SO$_2$ are strongly dependent on the source, which suggests that
they are related either to its evolutionary stage or to its initial conditions
as already suggested by chemical models (e.g.,  \citealt{1997ApJ...481..396C} for the dependence on the evolutionary stage,  \citealt{2004A&A...422..159W} for the dependence
on the initial conditions).

\section{Conclusion}

We observed the massive star forming region IRAS\,17233$-$3606 with the
APEX telescope in several molecular transitions of high diagnostic
value at a wavelength of 1~mm. Our observations reveal a bipolar
molecular outflow associated with strong emission at 4.5~$\mu$m, and
an exceptionally rich molecular spectrum from a position very close to
the centre of the outflow.  To better characterise the nature of this
source, we performed continuum cross-scans on the source at 0.9, 0.7
and 0.4~mm, and derived the flux density of the source over a beam
between 18 and 7.6$''$.

We modelled the molecular emission of the source using a LTE
approximation, and found that it is compatible with that of a hot core
of 150~K and a source size of 2.7$''$. The distance of the source, and
its bolometric luminosity, are not well known.  However, assuming that
the source is at a distance of 1~kpc, and that the bulk of the dust is
at 70~K, we found that its submillimetre continuum emission
corresponds to a mass in the range of 103--240~M$_\odot$.  We also
analysed the molecular emission associated with the bipolar outflow,
using the CO (3-2) transition and assuming that the emission is
optically thin and the gas is at 50~K. The physical parameters are
typical of those driven by high-mass young stellar objects. However,
follow-up observations in high energy and in low optically
depth transitions are needed for a better characterisation of the outflow.

If the estimate of the distance to the source is correct, IRAS\,17233$-$3606
is probably placed among the less studied hot cores around B-type
(proto)stars.  This seems to be confirmed by the parameters of the molecular
outflow, which put its powering source among the less luminous massive
sources. It is therefore an interesting target for future high-angular
resolution observations aimed at investigating the connection between
hot cores around objects of high mass and hot corinos found around
solar type protostars.  Moreover, given its close distance and its
location in sky, IRAS\,17233 will  easily be accessible in the near
future for high linear resolution studies with ALMA.

\begin{acknowledgements}
We would like to thank Leonardo Testi and Bernd Klein for helpful discussions. We appreciate the careful referee's report which helped improving the paper.
\end{acknowledgements}

\bibliographystyle{aa}
\bibliography{biblio_leu}

\begin{thebibliography}{63}
\expandafter\ifx\csname natexlab\endcsname\relax\def\natexlab#1{#1}\fi

\bibitem[{{Benjamin} {et~al.}(2003){Benjamin}, {Churchwell}, {Babler}, {Bania},
  {Clemens}, {Cohen}, {Dickey}, {Indebetouw}, {Jackson}, {Kobulnicky},
  {Lazarian}, {Marston}, {Mathis}, {Meade}, {Seager}, {Stolovy}, {Watson},
  {Whitney}, {Wolff}, \& {Wolfire}}]{2003PASP..115..953B}
{Benjamin}, R.~A., {Churchwell}, E., {Babler}, B.~L., {et~al.} 2003, \pasp,
  115, 953

\bibitem[{{Bergman}(1992)}]{1992PhDT.......252B}
{Bergman}, P. 1992, PhD thesis, , G{\"o}teborg, Sweden, (1992)

\bibitem[{{Beuther} {et~al.}(2002{\natexlab{a}}){Beuther}, {Schilke}, {Gueth},
  {McCaughrean}, {Andersen}, {Sridharan}, \& {Menten}}]{2002A&A...387..931B}
{Beuther}, H., {Schilke}, P., {Gueth}, F., {et~al.} 2002{\natexlab{a}}, \aap,
  387, 931

\bibitem[{{Beuther} {et~al.}(2002{\natexlab{b}}){Beuther}, {Schilke}, {Menten},
  {Motte}, {Sridharan}, \& {Wyrowski}}]{2002ApJ...566..945B}
{Beuther}, H., {Schilke}, P., {Menten}, K.~M., {et~al.} 2002{\natexlab{b}},
  \apj, 566, 945

\bibitem[{{Beuther} {et~al.}(2002{\natexlab{c}}){Beuther}, {Schilke},
  {Sridharan}, {Menten}, {Walmsley}, \& {Wyrowski}}]{2002A&A...383..892B}
{Beuther}, H., {Schilke}, P., {Sridharan}, T.~K., {et~al.} 2002{\natexlab{c}},
  \aap, 383, 892

\bibitem[{{Bontemps} {et~al.}(1996){Bontemps}, {Andre}, {Terebey}, \&
  {Cabrit}}]{1996A&A...311..858B}
{Bontemps}, S., {Andre}, P., {Terebey}, S., \& {Cabrit}, S. 1996, \aap, 311,
  858

\bibitem[{{Cabrit} \& {Bertout}(1990)}]{1990ApJ...348..530C}
{Cabrit}, S. \& {Bertout}, C. 1990, \apj, 348, 530

\bibitem[{{Carey} {et~al.}(2005){Carey}, {Noriega-Crespo}, {Price}, {Padgett},
  {Kraemer}, {Indebetouw}, {Mizuno}, {Ali}, {Berriman}, {Boulanger}, {Cutri},
  {Ingalls}, {Kuchar}, {Latter}, {Marleau}, {Miville-Deschenes}, {Molinari},
  {Rebull}, \& {Testi}}]{2005AAS...207.6333C}
{Carey}, S.~J., {Noriega-Crespo}, A., {Price}, S.~D., {et~al.} 2005, in
  Bulletin of the American Astronomical Society, Vol.~37, Bulletin of the
  American Astronomical Society, 1252

\bibitem[{{Charnley}(1997)}]{1997ApJ...481..396C}
{Charnley}, S.~B. 1997, \apj, 481, 396

\bibitem[{{Chin} {et~al.}(1996){Chin}, {Henkel}, {Whiteoak}, {Langer}, \&
  {Churchwell}}]{1996A&A...305..960C}
{Chin}, Y.-N., {Henkel}, C., {Whiteoak}, J.~B., {Langer}, N., \& {Churchwell},
  E.~B. 1996, \aap, 305, 960

\bibitem[{{Comito} {et~al.}(2005){Comito}, {Schilke}, {Phillips}, {Lis},
  {Motte}, \& {Mehringer}}]{2005ApJS..156..127C}
{Comito}, C., {Schilke}, P., {Phillips}, T.~G., {et~al.} 2005, \apjs, 156, 127

\bibitem[{{De Buizer} {et~al.}(2000){De Buizer}, {Pi{\~n}a}, \&
  {Telesco}}]{2000ApJS..130..437D}
{De Buizer}, J.~M., {Pi{\~n}a}, R.~K., \& {Telesco}, C.~M. 2000, \apjs, 130,
  437

\bibitem[{{Fa{\'u}ndez} {et~al.}(2004){Fa{\'u}ndez}, {Bronfman}, {Garay},
  {Chini}, {Nyman}, \& {May}}]{2004A&A...426...97F}
{Fa{\'u}ndez}, S., {Bronfman}, L., {Garay}, G., {et~al.} 2004, \aap, 426, 97

\bibitem[{{Fish} {et~al.}(2005){Fish}, {Reid}, {Argon}, \&
  {Zheng}}]{2005ApJS..160..220F}
{Fish}, V.~L., {Reid}, M.~J., {Argon}, A.~L., \& {Zheng}, X.-W. 2005, \apjs,
  160, 220

\bibitem[{{Fix} {et~al.}(1982){Fix}, {Mutel}, {Gaume}, \&
  {Claussen}}]{1982ApJ...259..657F}
{Fix}, J.~D., {Mutel}, R.~L., {Gaume}, R.~A., \& {Claussen}, M.~J. 1982, \apj,
  259, 657

\bibitem[{{Forster} \& {Caswell}(1989)}]{1989A&A...213..339F}
{Forster}, J.~R. \& {Caswell}, J.~L. 1989, \aap, 213, 339

\bibitem[{{Garay} {et~al.}(2007){Garay}, {Mardones}, {Bronfman}, {Brooks},
  {Rodr{\'{\i}}guez}, {G{\"u}sten}, {Nyman}, {Franco-Hern{\'a}ndez}, \&
  {Moran}}]{2007A&A...463..217G}
{Garay}, G., {Mardones}, D., {Bronfman}, L., {et~al.} 2007, \aap, 463, 217

\bibitem[{{G{\"u}sten} {et~al.}(2006){G{\"u}sten}, {Nyman}, {Schilke},
  {Menten}, {Cesarsky}, \& {Booth}}]{2006A&A...454L..13G}
{G{\"u}sten}, R., {Nyman}, L.~{\AA}., {Schilke}, P., {et~al.} 2006, \aap, 454,
  L13

\bibitem[{{Hatchell} {et~al.}(2000){Hatchell}, {Fuller}, {Millar}, {Thompson},
  \& {Macdonald}}]{2000A&A...357..637H}
{Hatchell}, J., {Fuller}, G.~A., {Millar}, T.~J., {Thompson}, M.~A., \&
  {Macdonald}, G.~H. 2000, \aap, 357, 637

\bibitem[{{Hatchell} {et~al.}(1998{\natexlab{a}}){Hatchell}, {Thompson},
  {Millar}, \& {MacDonald}}]{1998A&AS..133...29H}
{Hatchell}, J., {Thompson}, M.~A., {Millar}, T.~J., \& {MacDonald}, G.~H.
  1998{\natexlab{a}}, \aaps, 133, 29

\bibitem[{{Hatchell} {et~al.}(1998{\natexlab{b}}){Hatchell}, {Thompson},
  {Millar}, \& {MacDonald}}]{1998A&A...338..713H}
{Hatchell}, J., {Thompson}, M.~A., {Millar}, T.~J., \& {MacDonald}, G.~H.
  1998{\natexlab{b}}, \aap, 338, 713

\bibitem[{{Heyminck} {et~al.}(2006){Heyminck}, {Kasemann}, {G{\"u}sten}, {de
  Lange}, \& {Graf}}]{2006A&A...454L..21H}
{Heyminck}, S., {Kasemann}, C., {G{\"u}sten}, R., {de Lange}, G., \& {Graf},
  U.~U. 2006, \aap, 454, L21

\bibitem[{{Hieret} {et~al.}(2007){Hieret}, {Leurini}, {Menten}, {Schilke},
  {Thorwirth}, \& {Wyrowski}}]{2007arXiv0706.1643H}
{Hieret}, C., {Leurini}, S., {Menten}, K.~M., {et~al.} 2007, ArXiv e-prints,
  706

\bibitem[{{Hildebrand}(1983)}]{1983QJRAS..24..267H}
{Hildebrand}, R.~H. 1983, \qjras, 24, 267

\bibitem[{{Hughes} \& {MacLeod}(1993)}]{1993AJ....105.1495H}
{Hughes}, V.~A. \& {MacLeod}, G.~C. 1993, \aj, 105, 1495

\bibitem[{{Klein} {et~al.}(2006){Klein}, {Philipp}, {Kr{\"a}mer}, {Kasemann},
  {G{\"u}sten}, \& {Menten}}]{2006A&A...454L..29K}
{Klein}, B., {Philipp}, S.~D., {Kr{\"a}mer}, I., {et~al.} 2006, \aap, 454, L29

\bibitem[{{Kr\"{u}gel} \& {Walmsley}(1984)}]{1984A&A...130....5K}
{Kr\"{u}gel}, E. \& {Walmsley}, C.~M. 1984, \aap, 130, 5

\bibitem[{{Leurini} {et~al.}(2004){Leurini}, {Schilke}, {Menten}, {Flower},
  {Pottage}, \& {Xu}}]{2004A&A...422..573L}
{Leurini}, S., {Schilke}, P., {Menten}, K.~M., {et~al.} 2004, \aap, 422, 573

\bibitem[{{Leurini} {et~al.}(2006){Leurini}, {Schilke}, {Parise}, {Wyrowski},
  {G{\"u}sten}, \& {Philipp}}]{2006A&A...454L..83L}
{Leurini}, S., {Schilke}, P., {Parise}, B., {et~al.} 2006, \aap, 454, L83

\bibitem[{{Leurini} {et~al.}(2007){Leurini}, {Schilke}, {Wyrowski}, \&
  {Menten}}]{2007A&A...466..215L}
{Leurini}, S., {Schilke}, P., {Wyrowski}, F., \& {Menten}, K.~M. 2007, \aap,
  466, 215

\bibitem[{{M{\" u}ller} {et~al.}(2001){M{\" u}ller}, {Thorwirth}, {Roth}, \&
  {Winnewisser}}]{2001A&A...370L..49M}
{M{\" u}ller}, H.~S.~P., {Thorwirth}, S., {Roth}, D.~A., \& {Winnewisser}, G.
  2001, \aap, 370, L49

\bibitem[{{MacLeod} {et~al.}(1998){MacLeod}, {Scalise}, {Saedt}, {Galt}, \&
  {Gaylard}}]{1998AJ....116.1897M}
{MacLeod}, G.~C., {Scalise}, E.~J., {Saedt}, S., {Galt}, J.~A., \& {Gaylard},
  M.~J. 1998, \aj, 116, 1897

\bibitem[{{Miettinen} {et~al.}(2006){Miettinen}, {Harju}, {Haikala}, \&
  {Pomr{\'e}n}}]{2006A&A...460..721M}
{Miettinen}, O., {Harju}, J., {Haikala}, L.~K., \& {Pomr{\'e}n}, C. 2006, \aap,
  460, 721

\bibitem[{{Mitchell}(1984)}]{1984ApJ...287..665M}
{Mitchell}, G.~F. 1984, \apj, 287, 665

\bibitem[{{Molinari} {et~al.}(1996){Molinari}, {Brand}, {Cesaroni}, \&
  {Palla}}]{1996A&A...308..573M}
{Molinari}, S., {Brand}, J., {Cesaroni}, R., \& {Palla}, F. 1996, \aap, 308,
  573

\bibitem[{{Molinari} {et~al.}(2000){Molinari}, {Brand}, {Cesaroni}, \&
  {Palla}}]{2000A&A...355..617M}
{Molinari}, S., {Brand}, J., {Cesaroni}, R., \& {Palla}, F. 2000, \aap, 355,
  617

\bibitem[{{Molinari} {et~al.}(1998){Molinari}, {Brand}, {Cesaroni}, {Palla}, \&
  {Palumbo}}]{1998A&A...336..339M}
{Molinari}, S., {Brand}, J., {Cesaroni}, R., {Palla}, F., \& {Palumbo},
  G.~G.~C. 1998, \aap, 336, 339

\bibitem[{{Neckel}(1978)}]{1978A&A....69...51N}
{Neckel}, T. 1978, \aap, 69, 51

\bibitem[{{Noriega-Crespo} {et~al.}(2004){Noriega-Crespo}, {Morris}, {Marleau},
  {Carey}, {Boogert}, {van Dishoeck}, {Evans}, {Keene}, {Muzerolle},
  {Stapelfeldt}, {Pontoppidan}, {Lowrance}, {Allen}, \&
  {Bourke}}]{2004ApJS..154..352N}
{Noriega-Crespo}, A., {Morris}, P., {Marleau}, F.~R., {et~al.} 2004, \apjs,
  154, 352

\bibitem[{{Olmi} {et~al.}(1993){Olmi}, {Cesaroni}, \&
  {Walmsley}}]{1993A&A...276..489O}
{Olmi}, L., {Cesaroni}, R., \& {Walmsley}, C.~M. 1993, \aap, 276, 489

\bibitem[{{Ossenkopf} \& {Henning}(1994)}]{1994A&A...291..943O}
{Ossenkopf}, V. \& {Henning}, T. 1994, \aap, 291, 943

\bibitem[{{Osterloh} {et~al.}(1997){Osterloh}, {Henning}, \&
  {Launhardt}}]{1997ApJS..110...71O}
{Osterloh}, M., {Henning}, T., \& {Launhardt}, R. 1997, \apjs, 110, 71

\bibitem[{{Pickett} {et~al.}(1998){Pickett}, {Poynter}, {Cohen}, {Delitsky},
  {Pearson}, \& {Muller}}]{pickett_JMolSpectrosc_60_883_1998}
{Pickett}, H.~M., {Poynter}, I.~R.~L., {Cohen}, E.~A., {et~al.} 1998, Journal
  of Quantitative Spectroscopy and Radiative Transfer, 60, 883

\bibitem[{{Risacher} {et~al.}(2006){Risacher}, {Vassilev}, {Monje}, {Lapkin},
  {Belitsky}, {Pavolotsky}, {Pantaleev}, {Bergman}, {Ferm}, {Sundin},
  {Svensson}, {Fredrixon}, {Meledin}, {Gunnarsson}, {Hagstr{\"o}m},
  {Johansson}, {Olberg}, {Booth}, {Olofsson}, \& {Nyman}}]{2006A&A...454L..17R}
{Risacher}, C., {Vassilev}, V., {Monje}, R., {et~al.} 2006, \aap, 454, L17

\bibitem[{{Sandell}(2000)}]{2000A&A...358..242S}
{Sandell}, G. 2000, \aap, 358, 242

\bibitem[{{Schilke} {et~al.}(2006){Schilke}, {Comito}, {Thorwirth}, {Wyrowski},
  {Menten}, {G{\"u}sten}, {Bergman}, \& {Nyman}}]{2006A&A...454L..41S}
{Schilke}, P., {Comito}, C., {Thorwirth}, S., {et~al.} 2006, \aap, 454, L41

\bibitem[{{Schilke} {et~al.}(1997){Schilke}, {Groesbeck}, {Blake}, \&
  {Phillips}}]{1997ApJS..108..301S}
{Schilke}, P., {Groesbeck}, T.~D., {Blake}, G.~A., \& {Phillips}, T.~G. 1997,
  \apjs, 108, 301

\bibitem[{{Sch{\"o}ier} {et~al.}(2002){Sch{\"o}ier}, {J{\o}rgensen}, {van
  Dishoeck}, \& {Blake}}]{2002A&A...390.1001S}
{Sch{\"o}ier}, F.~L., {J{\o}rgensen}, J.~K., {van Dishoeck}, E.~F., \& {Blake},
  G.~A. 2002, \aap, 390, 1001

\bibitem[{{Sridharan} {et~al.}(2002){Sridharan}, {Beuther}, {Schilke},
  {Menten}, \& {Wyrowski}}]{2002ApJ...566..931S}
{Sridharan}, T.~K., {Beuther}, H., {Schilke}, P., {Menten}, K.~M., \&
  {Wyrowski}, F. 2002, \apj, 566, 931

\bibitem[{{van der Tak} {et~al.}(2003){van der Tak}, {Boonman}, {Braakman}, \&
  {van Dishoeck}}]{2003A&A...412..133V}
{van der Tak}, F.~F.~S., {Boonman}, A.~M.~S., {Braakman}, R., \& {van
  Dishoeck}, E.~F. 2003, \aap, 412, 133

\bibitem[{{Wakelam} {et~al.}(2004){Wakelam}, {Caselli}, {Ceccarelli}, {Herbst},
  \& {Castets}}]{2004A&A...422..159W}
{Wakelam}, V., {Caselli}, P., {Ceccarelli}, C., {Herbst}, E., \& {Castets}, A.
  2004, \aap, 422, 159

\bibitem[{{Walsh} {et~al.}(1998){Walsh}, {Burton}, {Hyland}, \&
  {Robinson}}]{1998MNRAS.301..640W}
{Walsh}, A.~J., {Burton}, M.~G., {Hyland}, A.~R., \& {Robinson}, G. 1998,
  \mnras, 301, 640

\bibitem[{{Walsh} {et~al.}(1999){Walsh}, {Burton}, {Hyland}, \&
  {Robinson}}]{1999MNRAS.309..905W}
{Walsh}, A.~J., {Burton}, M.~G., {Hyland}, A.~R., \& {Robinson}, G. 1999,
  \mnras, 309, 905

\bibitem[{{Walsh} {et~al.}(1997){Walsh}, {Hyland}, {Robinson}, \&
  {Burton}}]{1997MNRAS.291..261W}
{Walsh}, A.~J., {Hyland}, A.~R., {Robinson}, G., \& {Burton}, M.~G. 1997,
  \mnras, 291, 261

\bibitem[{{Williams} {et~al.}(2004){Williams}, {Fuller}, \&
  {Sridharan}}]{2004A&A...417..115W}
{Williams}, S.~J., {Fuller}, G.~A., \& {Sridharan}, T.~K. 2004, \aap, 417, 115

\bibitem[{{Williams} {et~al.}(2005){Williams}, {Fuller}, \&
  {Sridharan}}]{2005A&A...434..257W}
{Williams}, S.~J., {Fuller}, G.~A., \& {Sridharan}, T.~K. 2005, \aap, 434, 257

\bibitem[{{Wilner} {et~al.}(1995){Wilner}, {Welch}, \&
  {Forster}}]{1995ApJ...449L..73W}
{Wilner}, D.~J., {Welch}, W.~J., \& {Forster}, J.~R. 1995, \apjl, 449, L73

\bibitem[{{Wilson} \& {Rood}(1994)}]{1994ARA&A..32..191W}
{Wilson}, T.~L. \& {Rood}, R. 1994, \araa, 32, 191

\bibitem[{{Wright} {et~al.}(1996){Wright}, {Plambeck}, \&
  {Wilner}}]{1996ApJ...469..216W}
{Wright}, M.~C.~H., {Plambeck}, R.~L., \& {Wilner}, D.~J. 1996, \apj, 469, 216

\bibitem[{{Wyrowski} {et~al.}(2006){Wyrowski}, {Menten}, {Schilke},
  {Thorwirth}, {G{\"u}sten}, \& {Bergman}}]{2006A&A...454L..91W}
{Wyrowski}, F., {Menten}, K.~M., {Schilke}, P., {et~al.} 2006, \aap, 454, L91

\bibitem[{{Zapata} {et~al.}(2008){Zapata}, {Leurini}, {Menten}, {Schilke}, \&
  R.}]{zapata}
{Zapata}, L., {Leurini}, S., {Menten}, K.~M., {Schilke}, \& R., R. 2008, subm.

\bibitem[{{Zhang} {et~al.}(2005){Zhang}, {Hunter}, {Brand}, {Sridharan},
  {Cesaroni}, {Molinari}, {Wang}, \& {Kramer}}]{2005ApJ...625..864Z}
{Zhang}, Q., {Hunter}, T.~R., {Brand}, J., {et~al.} 2005, \apj, 625, 864

\bibitem[{{Zhang} {et~al.}(2001){Zhang}, {Hunter}, {Brand}, {Sridharan},
  {Molinari}, {Kramer}, \& {Cesaroni}}]{2001ApJ...552L.167Z}
{Zhang}, Q., {Hunter}, T.~R., {Brand}, J., {et~al.} 2001, \apjl, 552, L167

\end{thebibliography}

\Online
\begin{figure*}
\centering
\includegraphics[angle=-90,width=13cm]{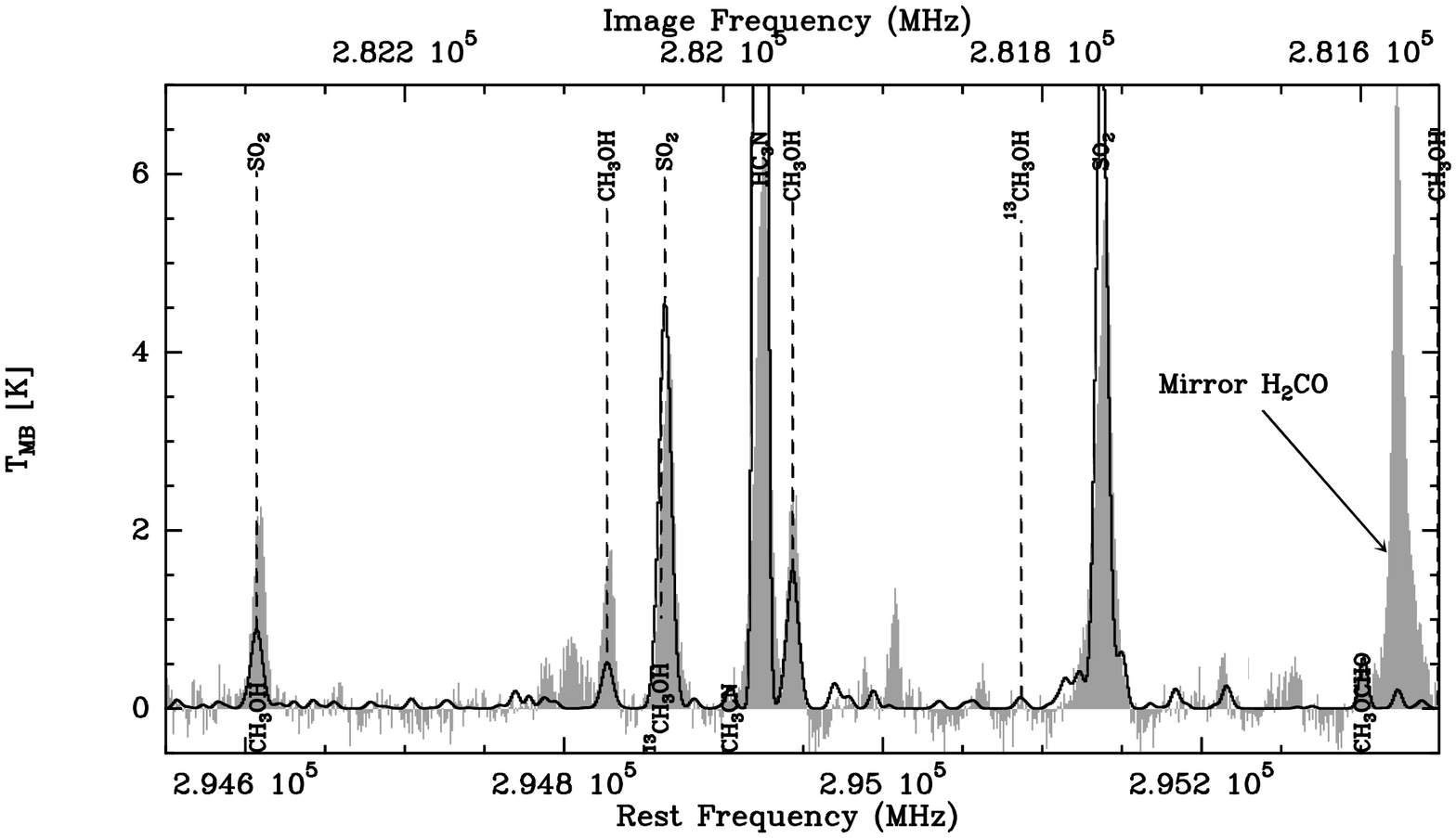}
\includegraphics[angle=-90,width=13cm]{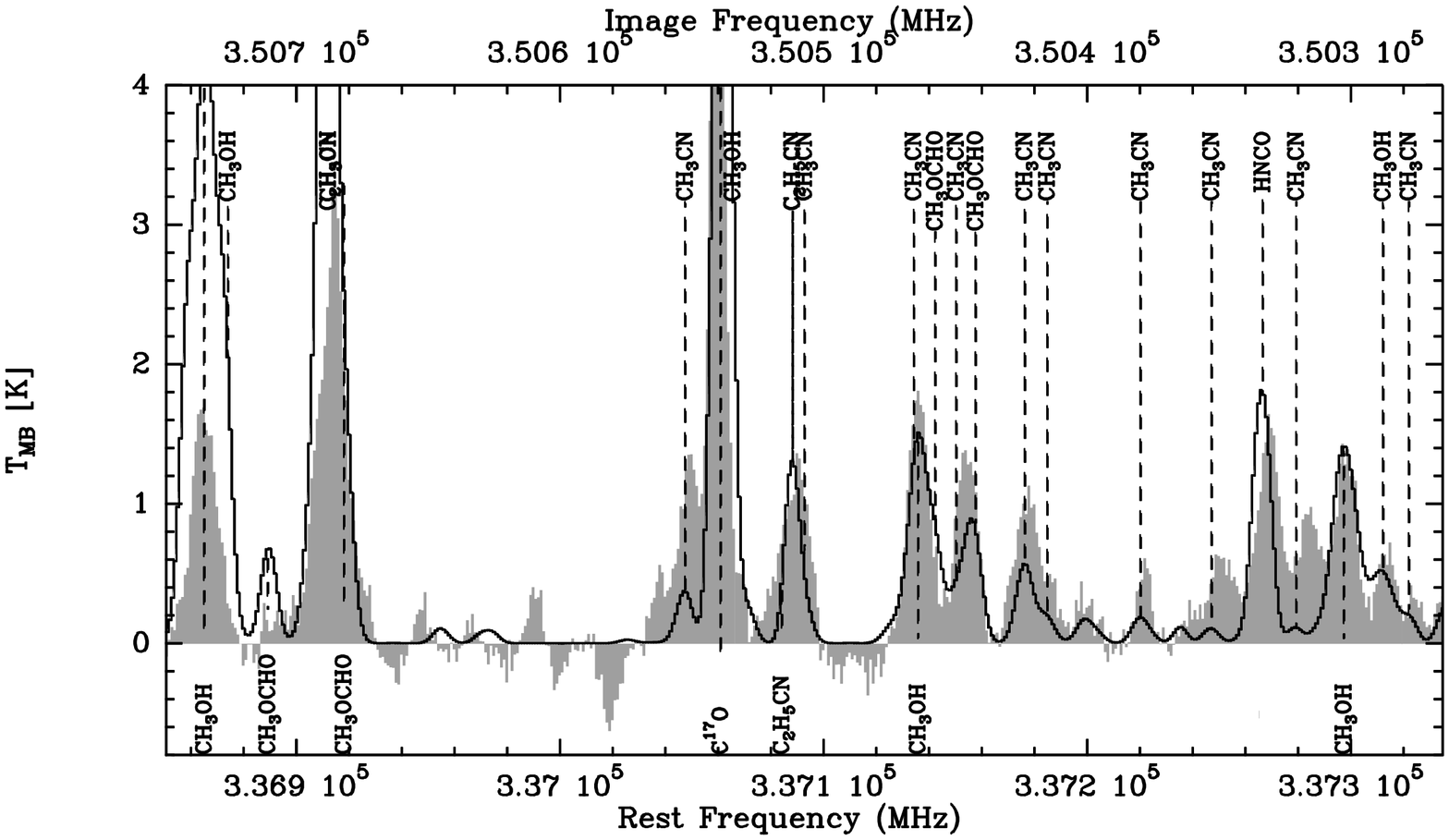}
\includegraphics[angle=-90,width=13cm]{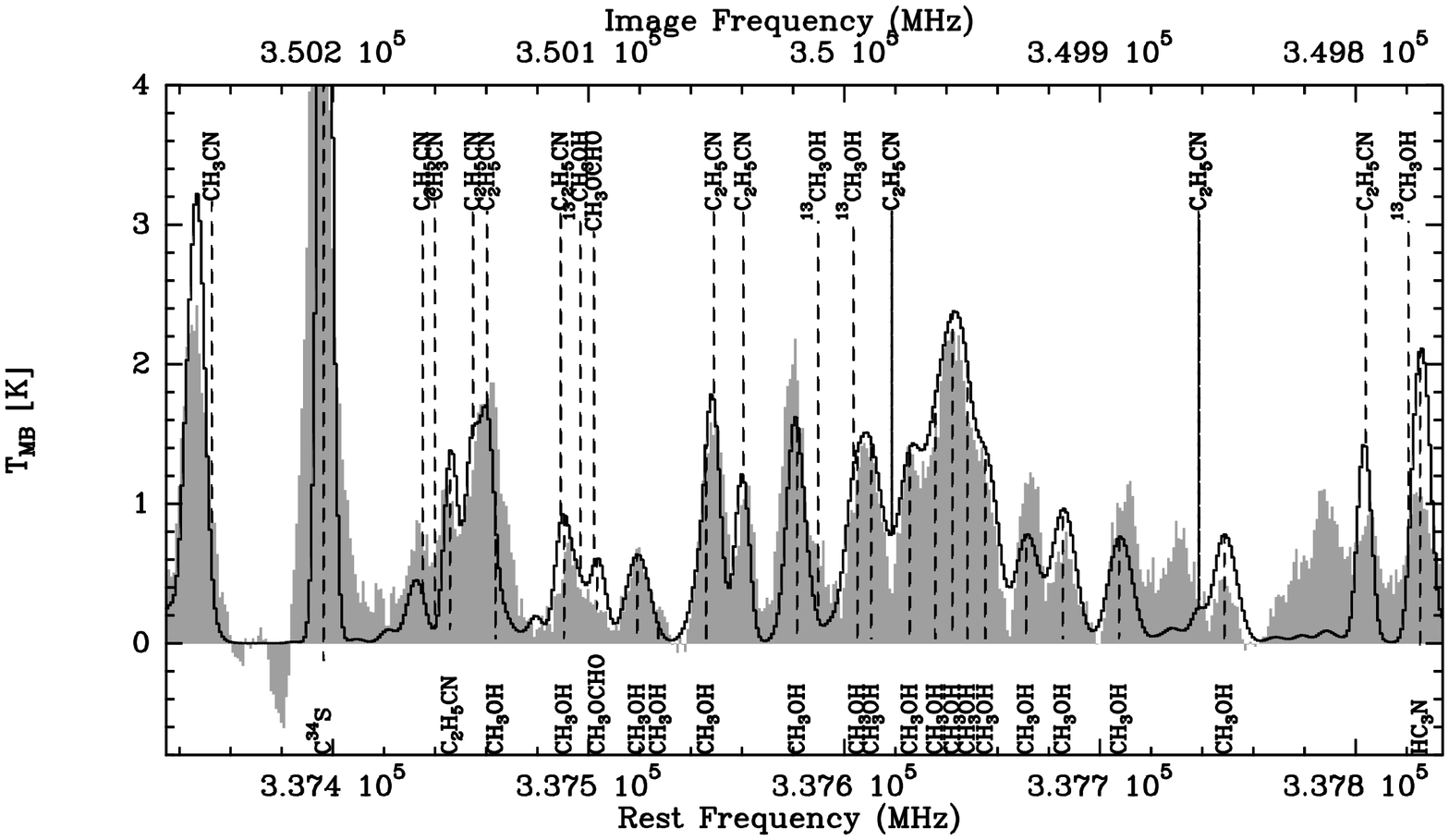}
\begin{center}
Fig.~\ref{otherspectra} -- Continued
\end{center}
\end{figure*}
\begin{figure*}
\centering
\includegraphics[angle=-90,width=13cm]{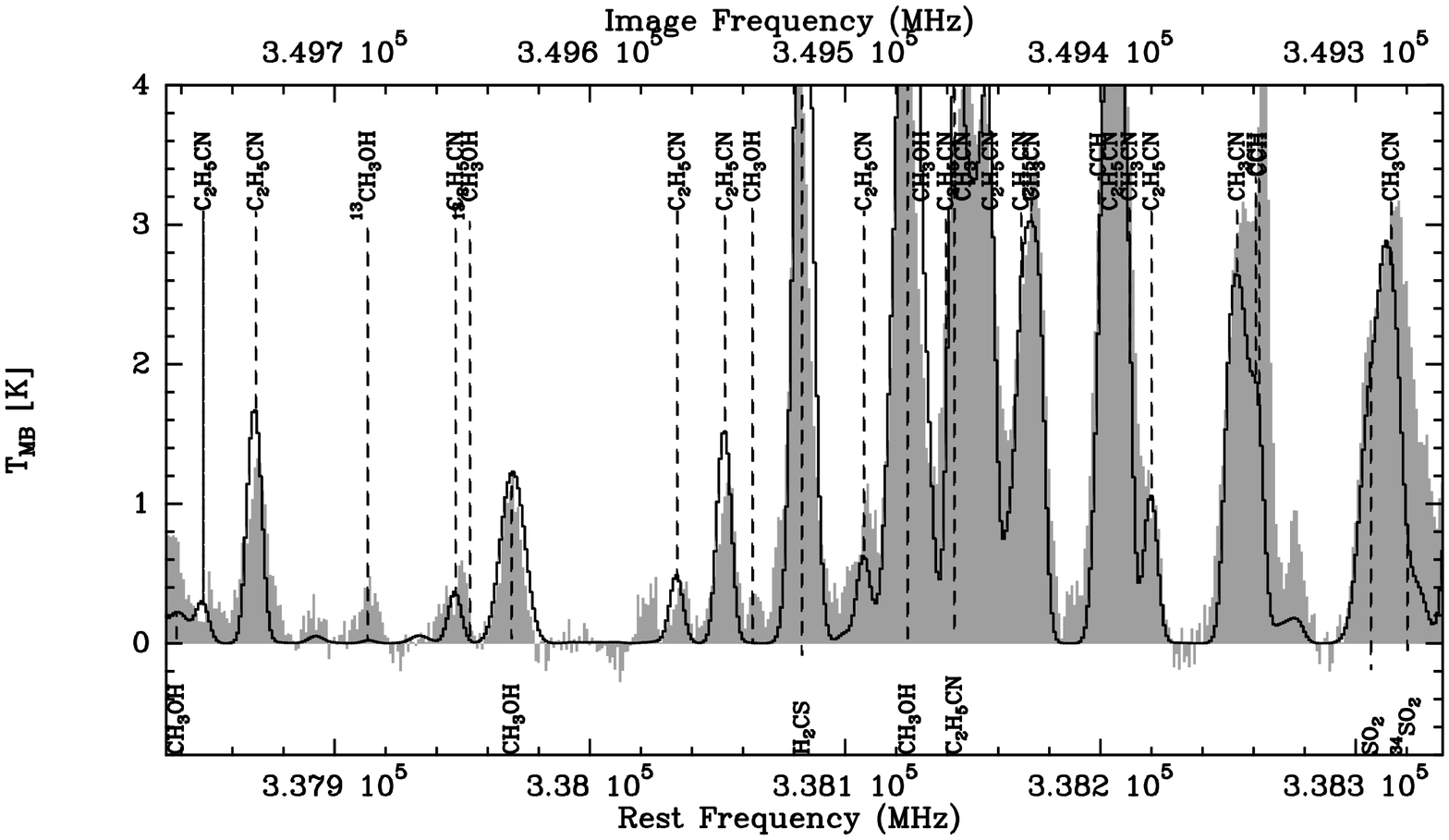}
\includegraphics[angle=-90,width=13cm]{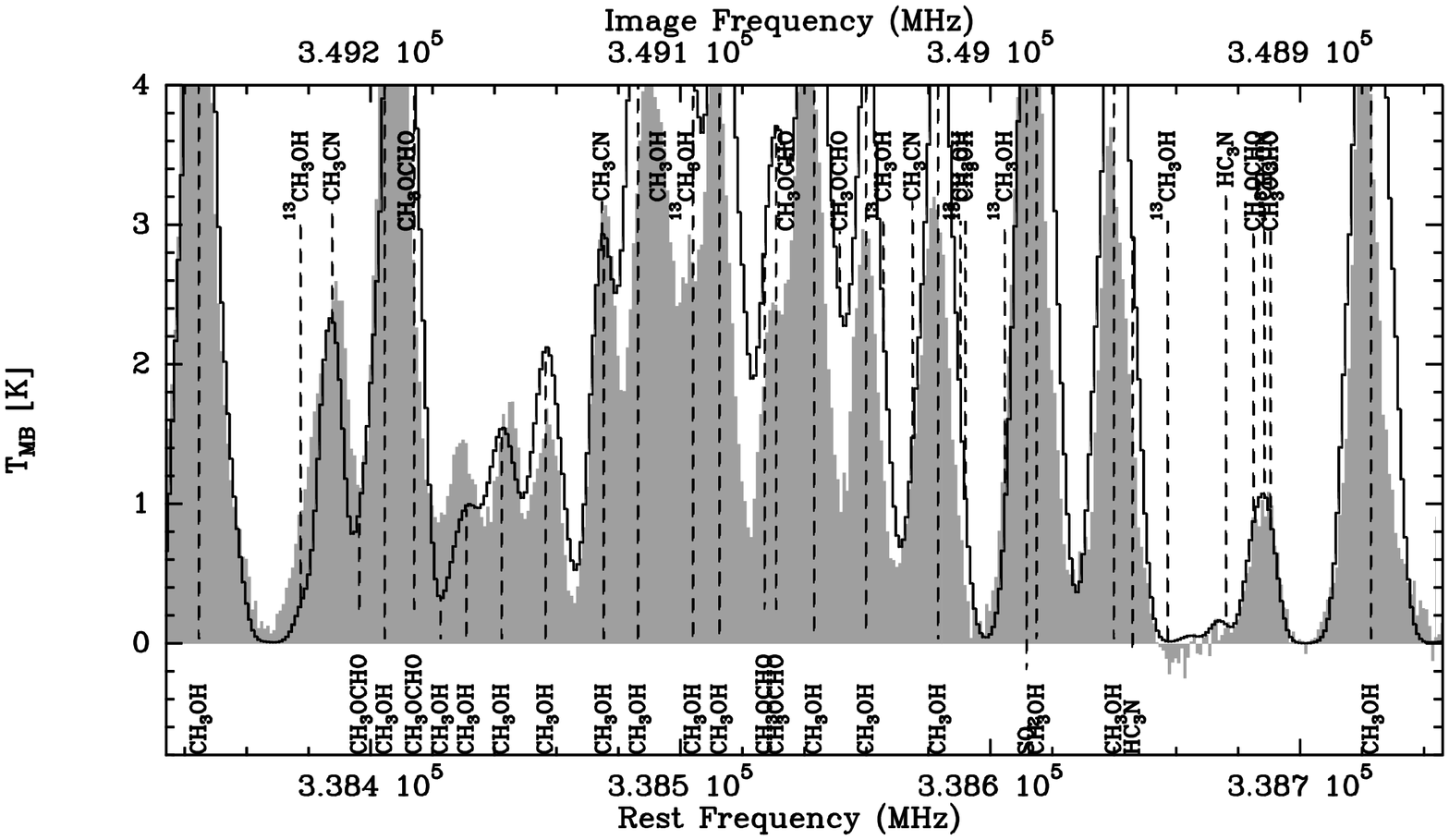}
\begin{center}
Fig.~\ref{otherspectra} -- Continued
\end{center}
\end{figure*}

\end{document}